\documentclass[fleqn,usenatbib]{mnras}

\usepackage[T1]{fontenc}

\DeclareRobustCommand{\VAN}[3]{#2}
\let\VANthebibliography\thebibliography
\def\thebibliography{\DeclareRobustCommand{\VAN}[3]{##3}\VANthebibliography}

\usepackage{blindtext}
\usepackage{cancel}
\usepackage{mathtools,mathalfa,amssymb,amsmath,amsfonts, bm}

\usepackage{amsthm}
\usepackage{graphicx}
\usepackage{multirow}
\usepackage[table, dvipsnames]{xcolor}
\usepackage{color}
\definecolor{292}{rgb}{0.3843, 0.6588, 0.8980}

\usepackage{soul}
\usepackage{orcidlink}
\usepackage{float}

\usepackage{appendix}
\usepackage{float}
\usepackage{subfigure}
\usepackage{xcolor}

\colorlet{green}{teal}
\graphicspath{{./images/Linear_Plots/}{/images/}{/images/new/}}

\newcommand{\nc}{\newcommand*} 
\nc{\figurewidth}{3.2in}
\nc{\xbar}{\bar{x}}
\nc{\rhoeq}{\rho_{\mathrm{eq}}}
\nc{\zeq}{z_{\mathrm{eq}}}
\nc{\tla}{\tilde{\lambda}}
\nc{\dt}{\delta}
\nc{\Dt}{\Delta}
\nc{\vj}{\vec{j}}
\nc{\vl}{\vec{l}}
\nc{\hx}{\hat{x}}
\nc{\hy}{\hat{y}}
\nc{\bj}{\bm{j}}
\nc{\mJ}{\mathcal{J}}
\nc{\mP}{\mathcal{P}}
\nc{\Msun}{M_\odot}
\nc{\app}{\approx}
\nc{\av}[1]{\langle #1 \rangle}
\nc{\eq}[1]{Eq.~\eqref{#1}}
\nc{\al}{\alpha}
\nc{\Xstar}{X_{\ast}}
\nc{\seq}{\sigma_{\mathrm{eq}}}
\nc{\fpbh}{f_{\mathrm{pbh}}}
\nc{\vth}{\vec{\theta}}
\nc{\vla}{\vec{\lambda}}
\nc{\vd}{\vec{d}}
\nc{\Mmin}{M_{\mathrm{min}}}
\nc{\rmd}{\mathrm{d}}
\nc{\mmin}{{m_{\mathrm{min}}}}
\nc{\mmax}{{m_{\mathrm{max}}}}
\nc{\mR}{\mathcal{R}}
\nc{\tmR}{\tilde{\mathcal{R}}}
\nc{\s}{\sigma}
\nc{\ogw}{\Omega_{\mathrm{GW}}}
\nc{\addref}{[\textcolor{red}{add ref}] }
\nc{\Om}{\Omega}
\nc{\gpcyr}{\mathrm{Gpc}^{-3}\,\mathrm{yr}^{-1}}
\nc{\Eq}[1]{Eq.~\eqref{#1}}
\nc{\Fig}[1]{Fig.~\ref{#1}}
\nc{\Table}[1]{Table~\ref{#1}}
\nc{\lvc}{LIGO/Virgo} % LIGO-VIRGO collaboration
\nc{\Sec}[1]{Sec.~\ref{#1}}
\nc{\eg}{\textit{e.g.~}}
\nc{\SNR}{\mathrm{SNR}}

\def\({\left(}
\def\){\right)}
\def\[{\left[}
\def\]{\right]}

\def\e{\begin{equation}}
\def\q{\end{equation}}
\def\m{\begin{eqnarray}}
\def\n{\end{eqnarray}}

\defcitealias{Goldreich:1979zz}{GT79}
\defcitealias{Goldreich:1980wa}{GT80}
\defcitealias{Goldreich:2013rma}{GS14}
\defcitealias{ward1988disk}{W88}

\title[Interfering Density Waves]{Mean-Motion Resonances With Interfering Density Waves }

\author[ H. Yang, Y. Li]{
Huan Yang $^{1,2,3}$\thanks{E-mail: hyangdoa@tsinghua.edu.cn}
Ya-Ping Li $^{4}$\thanks{E-mail: liyp@shao.ac.cn}
\\
$^{1}$Department of Astronomy, Tsinghua University, Beijing 100084, China \\
$^{2}$Perimeter Institute for Theoretical Physics, Waterloo, ON N2L2Y5, Canada\\
$^{3}$ University of Guelph, Guelph, Ontario N1G 2W1, Canada\\
$^4$ Shanghai Astronomical Observatory, Chinese Academy of Sciences, Shanghai 200030, People’s Republic of China\\
}

\date{Accepted XXX. Received YYY; in original form ZZZ}

\pubyear{2023}
\begin{document}
\label{firstpage}
\pagerange{\pageref{firstpage}--\pageref{lastpage}}
\maketitle

\begin{abstract}
In this work, we study the dynamics of two less massive objects moving around a central massive object, which are all embedded within a thin accretion disc. In addition to the gravitational interaction between these objects, the disc-object interaction is also crucial for describing the long-term dynamics of the multi-body system, especially in the regime of mean-motion resonances. We point out that near the resonance the density waves generated by  the two moving objects generally coherently interfere with each other, giving rise to extra angular momentum fluxes. The resulting backreaction on the objects is derived within the thin-disc scenario, which explicitly depends on the resonant angle and sensitively depends on the smoothing scheme used in the two-dimensional theory.  We have performed hydrodynamical simulations with planets embedded within a thin accretion disc and have found qualitatively agreement on the signatures of  interfering density waves by measuring the torques on the embedded objects, for the cases of $2:1$ and $3:2$ resonance.
By including in interference torque and the migration torques in the evolution of a pair of planets, we show that the chance of resonance trapping depends on the sign of the interference torque. For negative interference torques the pairs are more likely located at off-resonance regimes. The negative interference torques  may also explain the $1\%-2\%$ offset (for the period ratios) from the exact resonance values as observed in {\it Kepler} multi-planet systems. 

\end{abstract}

\begin{keywords}
accretion, accretion discs -- Planetary Systems -- gravitational waves -- planets and satellites: dynamical evolution and stability 
\end{keywords}

\section{Introduction}

Mean motion resonances (MMRs) generally arise for a system with two (or more) point masses orbiting around a common massive object. The mutual gravitational interaction between the two point masses gives rise to resonant dynamics of the resonant degree of freedom of this system when the period ratio is close to $j+k:j$ \citep{murray2000solar}, where both $j, k$ are integers. This general setting applies for various astrophysical systems at different scales, including satellites orbiting around planets \citep{Peale:1976bq}, planets orbiting around stars \citep{Petrovich:2012ev}, and stars/stellar-mass black holes orbiting around supermassive black holes \citep{Yang:2019iqa,Peng:2022hqa}. 
For example, it is known that the satellites Mimas and Tethys of Saturn are in a $4:2$ resonance with their mean motions being $n_{\rm M}:n_{\rm Te} =2.003139$ \citep{murray2000solar}.
%For example, it is  known that 
%the mean motion of Jupiter's satellite IO ($n_I$), Europa ($n_E$) and Ganymede $n_G$ satisfies $n_I-3n_E+2n_G=0$ to nine significant figures \citep{Peale:1976bq}. 
The {\it Kepler} mission has detected thousands of planets, some of which (although much less than expected) belong to multi-planet systems that exhibit resonant period ratios as well \citep{Lissauer:2012xe,Fabrycky:2012rh,Weiss2022,Huang2023,Dai2024}. On the other hand, the AGN (Active Galactic Nuclei) may capture stellar-mass black holes from the nuclear star cluster through the density wave generation \citep{tanaka2002three,tanaka2004three}. The embedded stellar-mass black holes  may be gravitationally captured into binaries within the AGN disc \citep{LiJ2023,DeLaurentiis2023,Rowan2023a,Rowan2024,Wang2024,Whitehead2024}, which subsequently become sources for ground-based gravitational wave detectors \citep[e.g.,][]{Li2021b,Li2022b,Dempsey2022,Kaaz2023,LiR2022a,LiR2023a,Lai2023}. They may also migrate towards the central massive black hole and eventually become extreme mass-ratio inspirals (EMRIs), which is expected to be an important or even dominant EMRI source for space-borne gravitational wave detectors \citep{Pan:2021ksp,Pan:2021oob,Pan:2021xhv}. If a pair of stellar-mass black holes are trapped into an MMR, they may migrate together towards the central massive black hole for certain period of time until the resonance locking breaks down \citep{Yang:2019iqa,Peng:2022hqa}. This possibly lead to subsequent EMRI events from the same galaxy with relatively short separations and/or gravitational environmental impacts on the EMRI waveform due to the tidal resonance effect \citep{Bonga:2019ycj,Pan:2023wau}. 

Starting from an initial state away from the MMR, a system may be captured into resonance due to additional dissipation mechanisms, i.e., migration torques from the disc. The probability of resonance capture depends on factors such as the migration direction, the initial orbital eccentricity, the masses of planets and central stars, etc \citep{murray2000solar}. In addition, \cite{Goldreich:2013rma} (hereafter \citetalias{Goldreich:2013rma}) showed that by incorporating migration torques into the long-term evolution of a pair of planets orbiting around a star, it is possible to explain that {\it Kepler} has observed much less than $50\%$ multi-planet systems trapped into MMRs \footnote{However, a refined treatment in \cite{deck2015migration} suggests that pairs of planets are more likely to be found near resonance following the formalism in \citetalias{Goldreich:2013rma}, but considering more general mass ratios.}. The period ratios for those residing near resonances are slightly larger than the exact resonances values, which is consistent with the requirement that resonance capture requires convergent migration. However, the average $1\%-2\%$ offset from the exact resonance values for the period ratios is difficult to explain within the framework of  \citetalias{Goldreich:2013rma}, so that it was conjectured additional mechanism is in operation to deduce the eccentricity and enhance the period ratio offset. There are debated arguments about whether tidal damping can account for the increased period ratio offset \citep{Lithwick:2012qt,Batygin:2012tu,Lee:2013pha}. It has also been suggested that 
dissipation of density waves near the planets may reverse the migration direction and/or increase the period ratio offset. It is however also worth to note that such mechanism is likely more relevant for massive planets with gap opening on the proto-planetary discs \citep{Baruteau:2013vv}. 

We notice that when there are multiple planets moving within a disc, the total density wave generated will be a superposition of waves contributed by each planet. In cases where planet orbital frequencies are not commensurate, the interference between different components of density waves only produce an oscillatory flux that averages to zero in time. So it is reasonable to expect to no secular effect associated with interfering density waves in this regime. However, when the planets are trapped in the resonance regime the density waves may stay in phase for an extended period of time so that their interference gives rise to additional angular momentum fluxes. The backreaction  should also modify the resonant dynamics of the pair of planets. 

In this work we explicitly compute the backreaction on the planets in resonant due to interfering density waves assuming a thin-disc scenario. 
%This additional torque modifies the planet orbital frequency and eccentricity in a timescale that is approximately $\mathcal{O}(10\%)$ of the migration timescale of the companion.
Using a two-dimensional disc perturbation theory along with a vertical smoothing scheme, we find that the backreaction torque  is a sinusoidal function of the resonant phase angle as analogous to the mutual gravitational interactions.  The torque is mainly produced in the regime where the location of a Lindblad resonance overlaps with the orbit of a planet. We find that the sign and magnitude of the torque, however, sensitively depends on the different smoothing schemes used in the two-dimensional theory. This means that the physical torque has to be computed in the three-dimensional setup. In order to further test the predictions of theory, we further carry out two-dimensional hydrodynamic simulations of these star$+$planets$+$disc systems using the FARGO3D code \citep{Benitez-Llambay2016}. 
%We have focused on a relatively simple  case that the outer planet is much more massive than the inner one, and the system is possible to be trapped in an MMR as the outer planet migrates inward with a rate faster than the inner planet.
By performing the simulations with a pair of planets and with individual planet, we can extract the additional torque due to the interference effect for the cases of $2:1$ and $3:2$ resonance. The result qualitatively agrees with the analytical in terms of the location of the peak torque density, the sign of the torque, with a factor of a few difference in the magnitude, which possibly comes from the approximation made in the analytical theory.

With the interference torque included in the orbital evolution of a pair of planets, we find that a positive interference torque (that tends to drive the inner planet outward) generally leads to higher chance of resonance locking, whereas a negative torque tends to produce more pairs of planets with period ratio away from the resonant values. This is interestingly consistent with the fact that  the majority of {\it Kepler}'s planet pairs are found away from resonant period ratios. We further examine the co-evolution of planet pairs with similar set-ups in \citetalias{Goldreich:2013rma}, in connection to {\it Kepler} observations. We find that for the same sets of planets and disc profiles, the interfering density wave terms are able to boost the period ratio offset by more than a factor of four, so that the observed offset level $\sim 1\%-2\%$ is much more compatible with the phenomenological evolution model discussed in \citetalias{Goldreich:2013rma}. Therefore the interfering density waves likely play an important role in  the morphology of astrophysical multi-planet systems.

This paper is organized as follows. In Sec.~\ref{sec:q} we perform an analytical calculation of the backreaction torque acting on planets due to the interfering density waves, under the thin-disc approximation. We further derive the consequent effect on the change rate of orbital eccentricity and frequency. In Sec.~\ref{sec:nume} we carry out hydrodynamical simulations of the multi-planet systems embedded within an accretion disc to test the theory of interfering density waves. In Sec.~\ref{sec:mod} we discuss the modified resonant dynamics due to the interfering density waves, with or without considering the migration torques.  In Sec.~\ref{sec:ob} we discuss the observational signatures of the interfering density waves in connection to the {\it Kepler} observations. We conclude in Sec.~\ref{sec:con}.

\section{Interfering Density Waves and their Backreaction}\label{sec:q}

Let us consider multiple  objects  moving within a thin disc, the gravitational fields of which excite density waves through the Lindblad resonance and the corotation resonance. Density waves carry away energy and angular momentum, which in turn lead to backreaction on the  objects as migration torques. These density waves generally have different frequencies as sourced by individual  objects, so that the interaction between one  object and density waves generated by other  objects should be oscillatory, i.e., no secular effect in the long-term evolution. On the other hand, it has been pointed out that density waves may be damped at co-orbital regions  of the  objects  due to dissipation of shocks \citep{Podlewska-Gaca:2011yrd}, so that these objects receive additional secular torques by dissipative interaction with density waves. This mechanism is efficient if  one or more objects have a partial gap opened to enhance the wave dissipation \citep{Baruteau:2013vv}. According to \citet{Podlewska-Gaca:2011yrd}, the resulting migration direction of planets may be reversed by this effect.

With the dissipative actions (e.g. density wave emission, tides) the multi-body systems may be locked into  MMRs, for which the period ratios are close to ratios of integers.
Notice that the analytical treatment presented in this section draws heavily from the formalism in \citet{Goldreich:1979zz} (hereafter \citetalias{Goldreich:1979zz}), \citet{Goldreich:1980wa} (hereafter \citetalias{Goldreich:1980wa}),\citet{ward1988disk} (hereafter \citetalias{ward1988disk}). The detailed derivation of some of the formulas from \citetalias{Goldreich:1979zz}, \citetalias{Goldreich:1980wa} and \citetalias{ward1988disk} are not repeated in this section, but instead cited with the corresponding equation numbers therein.
As a sample problem we consider an object ``A" moving on a {\it fixed} outer circular orbit and an object ``B" moving along an inner eccentric orbit.  The system is assumed to be locked in a $m:m-1$ resonance such that
\begin{align}
m \Omega_A = m \Omega_B -\kappa_B,
\end{align}
where $\Omega$ is the angular frequency and $\kappa$ is the epicyclic frequency.

The individual gravitational field for object A or B can be Fourier-decomposed as  ($s=A,B$)
\begin{align}\label{eq:phis}
& \varphi_{s}(r,\phi,t) \nonumber \\
&= \sum^\infty_{\ell_s=-\infty} \sum^\infty_{m_s=0} \varphi_{s,\ell_s,m_s} \cos \{ m_s \phi -[m_s \Omega_{s}+(\ell_s-m_s)\kappa_{s}]t\}\nonumber \\
& =  \sum^\infty_{\ell_s=-\infty} \sum^\infty_{m_s=0} \varphi_{s,\ell_s,m_s} \cos \{ m_s \phi -[\ell_s \lambda_s-(\ell_s-m_s)\varpi_s]\}
\end{align}
where we have used $\Omega_s =\dot{\lambda}_s$ and $\dot{\varpi}_s =\Omega_s-\kappa_s$ \citepalias{Goldreich:1980wa}. The  pattern frequency of the $\ell_s, m_s$ component is
\begin{align}
\Omega_{\ell_s,m_s} =\Omega_s + \frac{\ell_s-m_s}{m_s}\kappa_s \,.
\end{align}

The coefficients $\varphi_{s,\ell_s,m_s}$ as a function of the orbit parameters may be found in \citetalias{Goldreich:1980wa}. In the small eccentricity limit it is proportional to $e^{|\ell_s-m_s|}$. Therefore up to the linear order  in the eccentricity only the $|\ell_s-m_s|\le 1$ terms are relevant. Higher multiples matter if we go beyond the leading order effects. As object A is moving on a circular orbit, only $\ell_s=m_s$ term is nonzero for the expansion of its gravitational potential.  The second line of Eq.~\eqref{eq:phis} is more general than the first line as it does not assume the time dependent phase to be zero at $t=0$.

The gravitational field of the object A or B resonantly excites density waves in the disc through the Lindblad and corotation resonances, transferring part of the object's angular momentum to the disc. In particular,  the inner and outer Lindblad resonances are located at
 \begin{align}\label{eq:linc}
 m(\Omega-\Omega_{\ell_s,m_s}) =\pm \kappa, \quad m>0\,,
 \end{align}  
  where $m$ specifies  the particular Lindblad resonance between the planets, $\Omega, \kappa$ are orbital and epicyclic frequencies of the fluid, which are slightly different from orbital frequencies of the embedded compact objects. Notice that the location of the inner $m:m-1$ Lindblad resonance of object A should be close to the radius of object B, i.e.,
\begin{align}\label{eq:lind}
m (\Omega-\Omega_{m,m}) = m (\Omega-\Omega_{\rm A}) = \kappa\,.
\end{align}
For the corotation resonance the condition is
  \begin{align}
\Omega =\Omega_{\ell,m}\,.
\end{align}

The density waves produced by object $A,B$ can be decomposed into various harmonics with different azimuthal number and frequencies.
Denoting $\varphi^D_s$ ($s=A,B$) as the gravitational potential perturbation generated by density wave perturbations (e.g. see Eq.~(16) in \citetalias{Goldreich:1979zz}), the non-vanishing interference term produced by the harmonics appears when the system is locked in a resonance.
The relevant harmonics are given by (set the $m_A =m, \ell_A=m_A$ harmonics for $\varphi_A$ and $m_B=m, \ell_B=m_B-1$ harmonics for $\varphi_B$ in Eq.~\eqref{eq:phis}.)
 \begin{align}\label{eq:ha}
 \varphi^D_A(r) e^{i m \phi-i \omega t} &= \Phi_A e^{i\int^r_A ds k(s)} e^{i m \phi-i \omega t},\nonumber \\
 \varphi^D_B(r) e^{i m \phi-i \omega t} &= \Phi_B e^{i\int^r_B ds k(s)} e^{i m \phi-i \omega t} e^{i Q_0}
 \end{align}
with $\Phi_{A,B}$ being the amplitudes, $Q_0$ being a  constant related to the initial phase of object B's motion relative to object A, $k$ being the wave number, $\omega= m \Omega_A =m \Omega_B-\kappa_B$ and the integration bottom limit can be set as the location of the Lindblad resonance. Notice that other pairs of harmonics (e.g. the $l_B=m_B$ harmonic term here) that do not have the same azimuthal number and frequency can not produce a nonzero angular momentum flux through interference, after performing the angular and temporal average. On the other hand, according to the discussion  in \citetalias{Goldreich:1979zz} (Eq.~(30) therein), the angular momentum flux carried by a density wave is
\begin{align}
 \mathcal{F}_J = -{\rm sgn}(k) \frac{m r \Phi^2}{4 G} \left (1-\frac{c^2|k|}{\pi G \sigma} \right )
 \end{align}
where $\sigma$ is the surface density, $r$ is the radius of wave evaluation, $c^2=d P/d\sigma$ is the square of the sound speed and $\Phi$ is the amplitude of the total $\varphi^D$. Within the thin-disc approximation it can be shown that $ \mathcal{F}_J$ is independent of $r$. In \citetalias{Goldreich:1979zz} it is also shown that this angular momentum flux is equal to the migration torque (apart from the opposite sign) that backreacts on the object generating the density wave, which is expected from conservation of the total angular momentum.

With the superposition of density waves from object A and B, the total angular flux at a given radius receives beating terms that are proportional to the amplitudes of both waves. In particular, the beating between the harmonic components described by Eq.~\eqref{eq:ha}  leads to a nonzero (average-in-time) cross term
  \begin{align}
   \mathcal{F}_{J\times} & =  -{\rm sgn}(k) \frac{m r }{2 G} \left (1-\frac{c^2|k|}{\pi G \sigma} \right ){\rm Re}(\Phi_A \Phi_B e^{i (Q_0+C_{AB})}) \,.
   %& =-2{\rm sgn}(k) \sqrt{|\mathcal{F}_{Ja} \mathcal{F}_{Jb} |} \cos (Q+C_{ab})
  \end{align}
where $C_{AB}=\int^B_A ds\ k(s)$ is the additional phase factor coming from the wave propagation between  two Lindblad resonances.
This additional angular flux should correspond to additional migration torque acting on the  objects, but the angular momentum conservation alone cannot determine the fraction of torque exerted on each object. It requires specific analysis for the value of the torque on each massive object. In addition, in light of the Eq.~\eqref{eq:phis} it is obvious to see that the phase constant $Q_0$ should include the resonant angle $ m \lambda_A -(m-1)\lambda_B -\varpi_B$, where $\lambda$ and $\varpi$ are the mean longitude and longitude of pericenter respectively. Because of the dependence on the resonant angle, this additional torque (similar to the gravitational interactions) will introduce qualitatively different modification for the dynamics of the MMR, as compared to the type-I migration torques.

We shall explicit show that the nonzero flux cross term indeed gives rise to an additional torque between one planet and the density wave generated by the other planet. This torque generally averages to zero in time if the planet pair is not locked in resonance, but becomes nonzero and resonant-angle-dependent if the pair is in resonance. In Sec.~\ref{sec:ct} we show that for the MMR considered here, the dominant wave-planet coupling happens around the inner Lindblad resonance of planet A, which is close to the orbit of planet B as well. This observation is further verified by the numerical simulation in Sec.~\ref{sec:nume}.  In Sec.~\ref{sec:qr} we discuss the impact of this additional torque in the orbital evolution, i.e., how does it change the orbital frequency and eccentricity.

\subsection{Computing the torque}\label{sec:ct}

For the particular Lindblad resonance shown in Eq.~\eqref{eq:lind}, the relevant terms in the summation of gravitational harmonics in Eq.~\eqref{eq:phis} should have $m_A=\ell_A=m, m_B=m, \ell_B=m-1$. The corresponding $\varphi_{A,B}$ have the temporal and azimuthal dependence to allow a nonzero interference torque.

The density waves produced by the external gravitational potential $\varphi_{s}$ ($\varphi_{A,m,m}e^{i m \phi -i \omega t}$ and $\varphi_{B,m-1,m}e^{i m \phi -i \omega t}$) may be characterized by their associated density perturbation $\sigma_s(r)e^{i m \phi-i \omega t} $ ($\omega := m\Omega_{A}=m \Omega_B-\kappa_B$), velocity perturbation $(u_s \hat{e}_r +v_s \hat{e}_\phi)e^{i m \phi-i \omega t}$ and the gravitational perturbation $\varphi^D_s(r) e^{i m \phi-i \omega t}$ (e.g., see \citetalias{Goldreich:1979zz} for the wave equations governing these variables). In the spirit of  Eq. (A7) of \citetalias{Goldreich:1979zz}, the torque of the external potentials $\phi_A =\varphi_{A,m,m},\phi_B=\varphi_{B,m-1,m}$ acting on the disc is
\begin{align}
T = -m \pi \int ^\infty_0 dr r [\phi_A(r)+\phi_B(r)] {\rm Im}[\sigma_A(r)+\sigma_B(r)]\,.
\end{align}
  As a result, the backreaction of the density wave produced by object A on object B should be
  \begin{align}\label{eq:tb}
T_B  & =m \pi \int^\infty_0 dr r \phi_B(r) {\rm Im} \sigma_A(r) \nonumber \\
  & = -\pi m {\rm Im} \left \{ \int^\infty_0 dr \left [ \frac{m \phi_B \sigma v_A}{m \Omega -\omega} +i r \sigma u_A \frac{d}{dr} \left ( \frac{\phi_B}{m \Omega-\omega}\right )\right ] \right \}\,
  \end{align}
  where the second line is similar to Eq. (A8) of \citetalias{Goldreich:1979zz}. 
  
 \subsubsection{Lindblad resonances} \label{sec:lind}
  
  Near a Lindblad resonance we may use a different radial coordinate $x:=-1+r/r_L$ with $r_L$ being the radius of the Lindblad resonance. It can be obtained by requiring that (see Eq.~\eqref{eq:linc})
  \begin{align}
{D} = \kappa(r)^2- [m \Omega(r)-\omega]^2
  \end{align}
equals to zero at $r=r_L$. In this near zone of the Lindblad resonance, by solving the relevant wave equations the velocity perturbations can be obtained as (Appendix in \citetalias{Goldreich:1979zz}):
  \begin{align}\label{eq:ua}
  u_A =-\frac{\kappa}{r_L |\mathcal{D}|} \Psi_{A,\ell,m} \int^\infty_0 dt \,{\rm exp} \left [ i \left( t x -\frac{\alpha t^2}{2 \beta}+\frac{t^3}{3 \beta}\right ) \right ]
  \end{align}
  and
  \begin{align}
  v_A = i {\rm sgn}(\mathcal{D}) \frac{2 \Omega}{\kappa} u_A
  \end{align}
  where $\mathcal{D}$ is defined as $[r d D/dr]_{r_L}$, $\alpha$ is $(2\pi G \sigma r/c^2)_{r_L}{\rm sgn}(k)$, $\beta$ is $(r/c)^2_{r_L} \mathcal{D}$ and $\Psi_{s,\ell,m}$ is the source term in the wave equation:
  \begin{align}\label{eq:psislm}
\Psi_{s,\ell,m} = \left ( r\frac{d \varphi_{s,\ell,m}}{dr}+\frac{2 m \Omega}{m \Omega-\omega} \varphi_{s,\ell,m}\right )\,. 
  \end{align}
Here we have selected out the relevant harmonics with azimuthal number $m$ and  frequency $\omega$. Near the MMR, although the  frequency of relevant density waves generated by object A and B are both $\omega$, the wave variables may differ by a phase offset $Q = m \lambda_A -(m-1)\lambda_B -\varpi_B$. If we use object B as the phase reference, there is an additional factor of $e^{-i Q}$ multiplying the right hand side of Eq.~\eqref{eq:ua}. In addition, near the Lindblad resonance Eq.\eqref{eq:tb} can be further written as
\begin{align}\label{eq:tb2}
T_B = -\frac{\pi m \sigma r_L }{\kappa} {\rm sgn}(\mathcal{D}) \int^\infty_{-\infty} dx \Psi_{B,\ell,m} {\rm Re} (u_A ) f(x)
\end{align}
where the window function $f(x)$ is chosen such that $f=1$ near the resonance location and $f(x)\rightarrow 0$ for $|x|\rightarrow \infty$
to eliminate oscillatory contributions far away. 
%For a single object $s$, the Lindblad toque is given by }
%\begin{align}
%T_{\rm s} = \frac{\pi^2 m \sigma \Psi_{s,\ell,m}^2}{\mathcal{D}}\,
%\end{align}
%following the similar procedure in \citetalias{Goldreich:1979zz}. 

Let us consider the inner Lindblad resonance of object A, near which $\Psi_{B,\ell,m}$ may be written as (Eq.~$52$ in \citetalias{ward1988disk}) after taking into account the vertical average of the disc (so that the divergence is removed)
\begin{align}\label{eq:psibin}
\Psi_{B,m-1,m} =&-e_B \frac{G m M_B }{a_B \sqrt{\pi}} \left ( \frac{a_B \Omega}{c}\right ) [ \mathcal{F}_2(0,\xi) -4 \mathcal{F}_0(0,\xi) ] \\ \nonumber
&+\frac{4}{\pi} \frac{G M_B}{h \sqrt{\pi}} \sin f_c  
=\Psi_d+\Psi_c \sin f_c
\end{align}
where $M_B$ is the mass of object B, $e_B$ is its orbital eccentricity, $a_B$ is the semi-major axis, $h$ is the disc height and $\xi$ is defined as $\xi =m c/r\Omega$. Notice that for thin discs $c \sim \Omega h =\Omega r (h/r)$ so that $\xi \ll 1$ for low order m.  Here $f_c$ is related to $r$ as
\begin{align}
 \frac{r}{a_B}-1 =\gamma_0-1=-e_B \gamma_0 \cos f_c
\end{align}
The contribution from the $\Psi_c$ term becomes nonzero if $|1-\gamma_0|<e_B \gamma_0$, i.e., $r$ lies between aphelion and perihelion. The definitions of $\mathcal{F}$ functions are expressed as integrals of modified Bessel functions (Eq.~(26) of \citetalias{ward1988disk}):
\begin{align}
\mathcal{F}_0(\alpha_0,\xi) & =\pi^{-1} \int^\infty_{-\infty} e^{-(t/\xi)^2} K_0(\sqrt{\alpha_0^2+t^2}) dt\,,\nonumber \\
\mathcal{F}_2(\alpha_0,\xi) & =\pi^{-1} \int^\infty_{-\infty} e^{-(t/\xi)^2} \left \{ \frac{K_1(\sqrt{\alpha_0^2+t^2})}{\sqrt{\alpha_0^2+t^2}} - \frac{\alpha_0^2 K_2(\sqrt{\alpha_0^2+t^2})}{\alpha_0^2+t^2}\right \} dt\,.
\end{align}
In the limit that $\xi \ll 1$,  we have
\begin{align}
 \mathcal{F}_2(0,\xi) \approx \frac{2}{\sqrt{\pi}} \frac{1}{\xi}\,.
\end{align}
With $\Psi_{B,m-1,m}$ and $u_A$ the torque operating on object B can be evaluated following Eq.~\eqref{eq:tb2}. 
%following Eq.~\eqref{eq:tb2}. 
The torque $T_B$ can be explicitly written as 
\begin{align}\label{eq:tb2_re}
T_{\rm B, in} =& \frac{m \pi \sigma}{\mathcal{D}} \Psi_{A,m,m} \int^\infty_{-\infty} dx f(x)(\Psi_d+\Psi_c \sin f_c)  \nonumber \\
& \times {\rm Re} \left \{ \int^\infty_0 dt \,{\rm exp} \left [ i \left( t x -\frac{\alpha t^2}{2 \beta}+\frac{t^3}{3 \beta}-i Q\right ) \right ] \right \}
\end{align}

Notice that the $\Psi_d$ term  has no explicit $x$  dependence, so it can be moved outside of the integral.
By using the relation that (\citetalias{Goldreich:1979zz}, Appendix (a))
\begin{align}
&\int^\infty_{-\infty} dx \int^\infty_0 dt f(x) {\rm exp}\left [ i \left( t x -\frac{\alpha t^2}{2 \beta}+\frac{t^3}{3 \beta}\right ) \right ] \nonumber \\
&\approx 2\pi \int^\infty_0 dt \delta (t) {\rm exp}\left [ i \left(  -\frac{\alpha t^2}{2 \beta}+\frac{t^3}{3 \beta}\right ) \right ] =\pi\,,
\end{align}
we can identify the part of torque within $T_B$ that is associated with $\Psi_d$:

\begin{align}
  T_{\rm Bd} = \frac{\pi^2 m \sigma \Psi_{A,m,m} \Psi_{d}}{\mathcal{D}}\cos Q\,.
  \end{align}
On the other hand,  according to Eq.~\eqref{eq:psibin}, the $\Psi_c$ term in the integral of Eq.~\eqref{eq:tb2_re} has explicit dependence on $x$, which only makes nonzero contribution for the radius inside the orbital range $r \in [a_B(1-e_B),a_B(1+e_B)]$. This part of the integral is equal to ($dx = e_{\rm B} \sin f_c d f_c$)
\begin{align}\label{eq:integ}
T_{Bc} =&\frac{m \pi \sigma}{\mathcal{D}} \Psi_{A,m,m} \Psi_c \int^\infty_{-\infty} dx f(x) \sin f_c \Theta[e_B a_B-|r-a_B|] \nonumber \\
& \times {\rm Re} \left \{ \int^\infty_0 dt \,{\rm exp} \left [ i \left( t x -\frac{\alpha t^2}{2 \beta}+\frac{t^3}{3 \beta}-i Q\right ) \right ] \right \} \nonumber \\
\approx &\frac{m \pi \sigma}{\mathcal{D}} \Psi_{A,m,m} \Psi_c \int^\pi_{0} e_B \sin^2 f_c  f(x_0) df_c \nonumber \\
& \times {\rm Re} \left \{ \int^\infty_0 dt \,{\rm exp} \left [ i \left( t x_0 -\frac{\alpha t^2}{2 \beta}+\frac{t^3}{3 \beta}-i Q\right ) \right ] \right \} \nonumber \\
 =& \frac{m \pi^2 e_B \sigma}{2\mathcal{D}} \Psi_{A,m,m} \Psi_c 
 {\rm Re} \left [ \,W(\alpha ,\beta )e^{i Q}\right ]
%=-\frac{\pi m \sigma r_L \Psi_{c}}{\kappa} {\rm sgn}(\mathcal{D}) \int^\infty_{-\infty} dx {\rm Re} (\sin f_c u_a ) f(x) \,.
\end{align}
where $x_0=a_B/r_L-1 \sim \mathcal{O}(h^2/a_B^2)$ so that $f(x_0) \approx 1$ \footnote{A useful discussion regarding the difference between the fluid motion and the motion of the massive object can be found in Sec. V of \cite{Kocsis:2011dr}}, and we have defined a function $W(\alpha, \beta):=\int^\infty_0 dt \,{\rm exp} \left [ i \left(  -\frac{\alpha t^2}{2 \beta}+\frac{t^3}{3 \beta}\right ) \right ]$ for the integral in the last line. Notice that the integral in t becomes highly oscillatory when $t> {\rm Min}[\beta^{1/3},\sqrt{\beta/\alpha}]$, with $\alpha \sim r/(h Q_{\rm gas})$, $\beta \sim \mathcal{O}(r^2/h^2)$ ($Q_{\rm gas}$ is the Toomre parameter). We expect  both $\beta^{1/3} x_0 \ll 1$ and $\sqrt{\beta/\alpha} x_0 \ll 1$, so that the $ t x_0$ term in the integral may be removed. The resulting $T_{\rm B}$ (by including both the $T_{Bd}$ and $T_{Bc}$ components) is
\begin{align}\label{eq:bin}
  T_{\rm B, in} = \frac{\pi^2 m \sigma \Psi_{ A,m,m}}{\mathcal{D}}\left \{\Psi_d\cos Q  +\frac{e_B }{2}\Psi_c {\rm Re} \left [ \,W(\alpha ,\beta )e^{-i Q}\right ]\right \}\,.
\end{align}

%\begin{align}\label{eq:integ}
%&\int ^\infty_{-\infty} d x \sin f_c\int^\infty_0 dt \,{\rm exp} \left [ i \left( t x -\frac{\alpha t^2}{2 \beta}+\frac{t^3}{3 \beta}\right ) \right ] \nonumber \\
%& \approx \frac{\pi e}{2}\int^\infty_0 dt  {\rm exp} \left [ i \left( t x_0 -\frac{\alpha t^2}{2 \beta}+\frac{t^3}{3 \beta}\right ) \right ] \nonumber \\
%& \approx \frac{\pi e}{2}\sqrt{\pi/2}e^{-i \pi/4}\sqrt{\beta/\alpha}
%\end{align}
%where we have assumed that $e r/h <1$ to separate out the integration in $x$ and $t$. 

In general the two terms within the bracket together give rise to a sinusoidal term in the resonant angle $Q$.
%In the solar neighborhood the Toomre parameter is close to one \citepalias{Goldreich:1979zz}.
If we adopt the same approximation that the Toomre parameter $Q_{\rm gas}\sim \mathcal{O}(1)$ as the analysis in \citepalias{Goldreich:1979zz}, we expect $\beta^{1/3} \gg \sqrt{\beta/\alpha}$ and $W(\alpha,\beta) \approx \sqrt{\pi \beta/2 \alpha}e^{-i \pi/4}$.
%For more general scenarios the above integral should be carried out numerically (see the discussion in Sec.~\ref{sec:nume}).
% As $\alpha \sim r/h /Q_{\rm gas}, \beta \sim \mathcal{O}(r^2/h^2)$, the relevant range of integration for $t$ is $\sim \sqrt{\beta/\alpha} \sim \sqrt{r/h}$ so that $x_0 t \ll 1$. The resulting $T_{\rm B}$ (by including the $T_{Bd}$ component) is
In this limit the inner Lindblad torque becomes
\begin{align}\label{eq:bin2}
  T_{\rm B, in} = \frac{\pi^2 m \sigma \Psi_{ A,m,m}}{\mathcal{D}}\left [\Psi_d\cos Q  +\frac{e_B }{2}\sqrt{\pi \beta/2\alpha}\Psi_c \cos(Q+\pi/4)\right ]\,.
\end{align}
On the other hand, in the limit that the disc self-gravity is negligible, we expect that $\beta^{1/3} \ll \sqrt{\beta/\alpha}$. In this limit we have 
\begin{align}
W(\alpha,\beta) \approx \left \{\frac{i \Gamma(1/3)}{2\times 3^{2/3}}+\frac{2 \pi}{2\times 3^{2/3}\Gamma[2/3]} \right \} \beta^{1/3}\,.
\end{align}
Plugging this expression into Eq.~\eqref{eq:bin} we can obtain the corresponding torque expression
\begin{align}\label{eq:bin3}
  T_{\rm B, in} \approx \frac{\pi^2 m \sigma \Psi_{ A,m,m}}{\mathcal{D}}\left [\Psi_d\cos Q  +0.64e_B \beta^{1/3}\Psi_c \cos(Q-0.52)\right ]\,.
\end{align}

For the outer Lindblad resonance of object A, $\Psi_{B,m-1,m}$ is approximately constant in the relevant resonance range, so that
\begin{align}
  T_{\rm B,out} = \frac{\pi^2 m \sigma \Psi_{A,m,m} \Psi_{B,m-1,m}}{\mathcal{D}}\cos Q\,.
  \end{align}
The expression is  regular within the thin disc approximation so that the vertical average is not needed. Its value is given by
\begin{align}
\Psi_{B,m-1,m} = -e_B\frac{G M_B}{2 a_{\rm B}} \left [ \gamma^2 \frac{d^2 b^m_{1/2}}{d \gamma^2} -4 m \gamma \frac{d b^m_{1/2}}{d \gamma}+4 m^2 b^m_{1/2}\right ]
\end{align}
where $b^m_{1/2}(\gamma)$ are the Laplace coefficients 
\begin{align}
b^m_{1/2}(\gamma) \equiv \frac{2}{\pi} \int^\pi_0 \frac{\cos m \phi d \phi}{(1-2 \gamma \cos \phi+\gamma^2)^{1/2}}
\end{align}
and the above expression should be evaluated at $\gamma_0 = [(m+1)/(m-1)]^{2/3}$ in the case of a Keplerian disc. As we compare $\Psi_{B,m-1,m}$ evaluated at the inner and outer Lindblad resonance of object B, we find that $\Psi_{B,m-1,m}$ of the inner Lindblad (e.g. Eq.~\eqref{eq:psibin}) is larger than the one for outer resonance by a factor $\sim 1/\xi$. So the contribution from outer Lindblad resonance can be neglected for thin discs.

In summary, the density wave generated by planet A backreacts on planet B mainly in the neighborhood of the inner Lindblad resonance of planet A, which is also close to the orbit of planet B. The backreaction torque is given by Eq.~\eqref{eq:tb2} or Eq.~\eqref{eq:bin}, with the latter being derived in the vertical average approximation of the disc. Notice that when the planet pair is off resonance, there is still instantaneous torque from the wave-planet interaction. However, they generally average to zero in time as the frequency of the wave $m \Omega_A$ mismatches with the relevant harmonic frequency of the planet potential $m \Omega_B-\kappa_B$.

\subsubsection{Corotation resonance}

For a single object $s$, the torque due to corotation resonnace is 
\begin{align}
T_{\rm s, co} = -\frac{m \pi^2}{2} \left [ \left ( \frac{d \Omega}{ d r}\right )^{-1}\frac{d}{d r} \left ( \frac{\sigma}{B_0}\right )(\varphi_{s,\ell,m})^2\right ]
\end{align}
with $B_0 := \Omega+r/2 d \Omega/dr$. In analogy with the case of Lindblad resonances, the interaction between density wave generated by object A and object B produces a corotation torque:
\begin{align}
T_{\rm B, co} = -\frac{m \pi^2}{2} \left [ \left ( \frac{d \Omega}{ d r}\right )^{-1}\frac{d}{d r} \left ( \frac{\sigma}{B_0}\right )\varphi_{A,m,m}\varphi_{B,m-1,m} \cos Q\right ]_{r_C}
\end{align}
evaluated at the location of corotation resonance $r_C$. Notice that $\varphi_{A,m,m}$ here needs to be averaged over the vertical scale of the disc, which gives rise to \citepalias{ward1988disk}
\begin{align}
\langle \varphi_{A,m,m} \rangle = -\frac{2 G M_A \mathcal{F}_0(0,\xi)}{ r \sqrt{\pi}} \xi^{-1}\,.
\end{align}
Since the factor $\mathcal{F}_0(0,\xi) \propto \xi$ for small $\xi$, it means that $\langle \varphi_{A,m,m} \rangle$ has no explicit $\xi$ dependence. On the other hand, we have
\begin{align}
\varphi_{B,m-1,m} = -e_{\rm B} \frac{G M_B}{2 a_{\rm B}} \left [\gamma \frac{d b^m_{1/2}}{d \gamma}+(1-2m) b^m_{1/2} \right ]\,.
\end{align}
The resulting $T_{\rm B, co}$ is approximately $\mathcal{O}(\xi^{2} ) \sim\mathcal{O}(h^2/r^2 )$ times smaller than $T_{\rm B, in}$, so we shall neglect this piece in the rest of the discussions.

\subsubsection{Alternative smoothing scheme}\label{sec:alt}

\begin{figure}
\centering
\includegraphics[clip=true, width=0.45\textwidth]{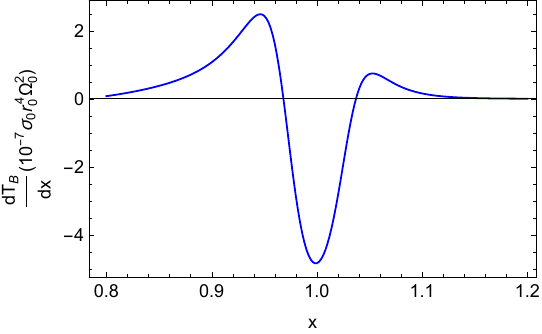}
\caption{Torque density assuming a pair of planets trapped in a $2:1$ resonance, with the resonance angle $Q$ being zero and the same binary configuration used as the $e_B=0.1$ setup in Sec.~\ref{sec:nume}. } 
	\label{fig:torque_21}
\end{figure}

\begin{figure}
\centering
\includegraphics[clip=true, width=0.45\textwidth]{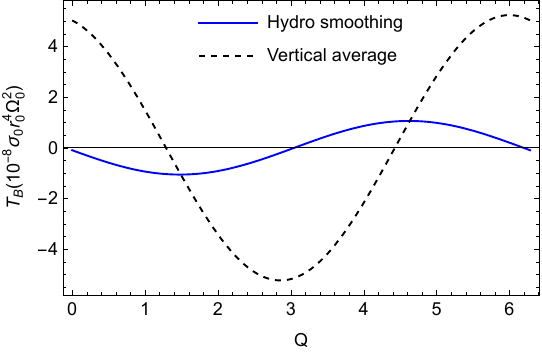}
\caption{Comparison of the torque as a function of $Q$ for two different smoothing schemes. The orbital configuration is the same as the $e_B=0.1$ setup in Sec.~\ref{sec:nume}.  } 
	\label{fig:torque_com}
\end{figure}

In previous sections we have discussed a vertical averaging scheme introduced in \citetalias{ward1988disk} to regularize the torque generated around the co-orbital regime. In numerical simulation an alternative smoothing scheme is often implemented by modifying the gravitational potential produced by planets, i.e.
\begin{eqnarray}
    \Phi_{s} =   -\frac{G \mu_{s}M}{(\left|\bm{r}_{{\rm p},s}-\bm{r}\right|^2+\epsilon^2)^{1/2}}
    + \mu_{s} \Omega_{{\rm p},s}^{2} \bm{r}_{{\rm p},s} \cdot \bm{r}
    ,
    \label{eq:potential}
\end{eqnarray}
where $s=A,B$ for each planet, the first term is the direct potential from the  planet, and the second term is the indirect potential arising from the  choice of our coordinate system. The smoothing parameter $\epsilon$ prevents the numerical divergence of the potential in the co-orbital regime.

With the modified potential one can compute $\Psi_{s,\ell,m}$ according to Eq.~\eqref{eq:psislm} and the Lindblad torque  according to Eq.~\eqref{eq:tb2}. Notice that Eq.~\eqref{eq:tb2} represents the total torque across the resonance, where its integrand represents the torque density. In Fig.~\ref{fig:torque_21} we show the torque for a $2:1$ resonance case, with $\epsilon=0.6 h$ and other parameters consistent with the numerical simulation discussed in Sec.~\ref{sec:nume}. We indeed find torque density centred around the location of the inner Lindblad resonance. 
%There are however two important differences between the linear theory presented here and the numerical simulation in Sec.~\ref{sec:nume}. 
%First, the linear theory assume no viscosity in the disc whereas the numerical disc assumes significant $\alpha$ viscosity. 
%Second, the disc in the linear theory assumes a polytropic equation of state where the numerical disc assumes locally isothermal profile. 
%Therefore the torque density will not be the same but should be qualitatively similar. 

In Fig.~\ref{fig:torque_com} we compare the interference torques computed using two different smoothing schemes. The smoothing scheme consistent with hydrodynamical simulations is based on Eq.~\eqref{eq:tb2} and Eq.~\eqref{eq:psislm}, whereas the torque associated with the vertical average scheme discussed in Sec.~\ref{sec:lind} uses Eq.~\eqref{eq:bin}. The two predictions are different, which must come from different regularization methods used for contribution of disc materials in the corotation regime. Therefore an accurate quantification of the interference torque should include a three dimensional  calculation. 
 
\subsection{Change rate of orbital quantities}\label{sec:qr}

The additional torque due to interfering density waves affects the orbital motion of object B. In this section we evaluate the corresponding $e_{\rm B}$ and $\dot{n}_{\rm B}$, where $n_{\rm B}$ is the mean motion.
We also assume a Keplerian disc profile to simplify the calculations, which may also serve as order-of-magnitude estimation for more general scenarios.
For a Keplerian disc we expect
\begin{align}
\mathcal{D} =3(m-1)\Omega^2_B\,.
\end{align}
and 
\begin{align}
\frac{\beta}{\alpha} = \frac{3(m-1)}M{2\pi \sigma a_{\rm B}^2}\,.
\end{align}
To evaluate the Lindblad torque in Eq.~\eqref{eq:bin} $\Psi_{A,m,m}$ is also required:
\begin{align}
\Psi_{A,m,m} = -\frac{G M_A}{r_A}\left [ \gamma \frac{d b^m_{1/2}}{d \gamma}+2 m b^m_{1/2} \right ]_\gamma
\end{align}
where $\gamma$ should be set as $\gamma=(1-1/m)^{2/3}$. For a particular $m-1:m$ MMR between object B and A, only the density waves with the same $m$ are relevant for obtaining the ``resonant torque".  Let us consider the case with $m=2$ which will be further discussed in Sec. \ref{sec:mod}:
\begin{align}
\Psi_{A,2,2} = -2.4\frac{G M_A}{r_A}\,.
\end{align}
For object B 's orbit, the change rate of eccentricity due to the additional torque is \citepalias{ward1988disk}
\begin{align}
\frac{1}{e_B}\frac{d e_B}{d t}  = &\left [ \Omega_{m-1,m} -\Omega_B-2e_B^2 \Omega_B \left ( 1+\frac{d \log \kappa}{d \log r}\right )_{r=a}\right ] \nonumber \\
&\times \frac{T_{\rm B,in}}{M_B e_B^2 a_B^2\kappa^2} 
\end{align}
where we only use the dominant contribution from $T_{\rm B,in}$. 
In the case of a Keplerian disc and that  $e_{\rm B} \ll 1$, the $e_{\rm B}^2$ term within the square bracket can be neglected and the $\dot{e}_{\rm B}$ 
%(for $m=2$) 
simplifies to
\begin{align}\label{eq:dtau0}
\frac{d e_B}{d t}  = & \frac{-1}{m} \frac{T_{\rm B, in}}{M_B e_B a_B^2 \Omega_B}
% \approx  \,3 \left( \frac{m_A}{M}\right )  \frac{\sigma a_B^2}{M} \left( \frac{a_B}{h}\right)^2 \Omega_B \nonumber \\
%\times & \left [ 1.04 \left( \frac{h}{a_B}\right )^{1/3}\cos(Q-0.52)-\cos Q \right ] \nonumber \\
 :=   - \frac{\cos (Q-Q_0)}{\tau_0}
\end{align}
so that the sign of $\dot{e}_{\rm B}$ depends on the relative phase $Q_0$ encoded in  the $\cos (Q-Q_0)$  term. 
In order to compute $n_{\rm B}$, we use the energy balance equation that 
\begin{align}
\frac{d E}{d t} =\Omega_B T_{\rm B, in}
\end{align}
which naturally leads to
\begin{align}
\dot{n}_{\rm B}/n_{\rm B} & =-3 (G M)^{-2/3} n^{1/3} M_B^{-1} T_{\rm B, in} \nonumber \\
&:= - \frac{3 m e_{\rm B}}\cos (Q-Q_0){\tau_0}\,.
\end{align}
%Notice that the ratio between $\tau_+$ and $\tau_0$ is
%\begin{align}
%\frac{\tau_0}{\tau_+} =\sqrt{\frac{3  Q_{\rm gas}}{4} \frac{h}{a_B}}\,.
%\end{align}
In general $\tau_0$ and $Q_0$ should be obtained and calibrated with three-dimensional hydro-simulations. In Sec.~\ref{sec:mod} we investigate the simplest scenarios with $Q_0$ being 0 or $\pi$ (or equivalently $\tau_0$ being either postive or negative) where the resonant dynamics shows distinctive features.
%For thin discs with $h/a \ll 1$, 
%assuming that the Toomree parameter satisfies $Q_{\rm gas} \sim \mathcal{O}(1)$ ,
%the $\tau_+$ term is subdominant comparing to the $\tau_0$ term so that it may be neglected in the equation of motion. This is the assumption we adopt in the discussion in Sec.~\ref{sec:mod}. However, in more general settings  $W(\alpha,\beta)$ needs to be evaluated numerically to obtain the torque and the corresponding decay rate.
%the $\tau_+$ term may not be negligible.

%\begin{figure*}
%\centering
%\includegraphics[clip=true, width=0.45\textwidth]{aorbit_sim.pdf}
%\includegraphics[clip=true, width=0.45\textwidth]{ecc_sim.pdf}
%\caption{Left panels: the semi-major axis (upper left) and orbital period ratio (lower left) evolution for a pair of planets with $\mu_{B}=10^{-5}$ and $\mu_{A}=3\times10^{-4}$. The planet pair is captured into 2:1 MMR around 4000 orbits. 
%Right panel: the orbital eccentricity evolution for the planet pair. The two vertical dotted lines show 1000 and 5000 orbits just before and after the 2:1 MMR.
%Note that the times in this figure are all measured in unit of the orbital period at $r=r_{0}$.} 
%\label{fig:orbit_sim}
%\end{figure*}

\section{Hydrodynamical simulations}\label{sec:nume}

\subsection{Numerical Setups}

In order to further demonstrate the effect of interfering density waves beyond the analytical calculations, a few hydrodynamical simulations are carried out to confirm the effect of interfering density waves on the dynamics of the MMR pair.
We use the FARGO3D code \citep{Benitez-Llambay2016} to simulate the gravitational interaction of a pair of (or a single) planet with a 2D thin disc.
The disc aspect ratio is initialized as $h/r=0.05$ and is constant over the entire disc. As a result, we set a locally isothermal equation of state with a temperature profile of a power-law index of $-1.0$. The surface density is set as $\sigma=\sigma_{0}(r/r_0)^{-0.5}$ initially, where $\sigma_{0}=3\times10^{-5}M/r_{0}^{2}$ is the surface density at $r_{0}$, where $M$ is the stellar mass, $r_{0}$ is the typical length scale of the disc, which can be scaled freely to compare with observations. Such a low surface ensures that the disc self-gravity can be safely ignored in simulations.
We also assume an $\alpha$-prescription for the gas kinematic viscosity $\nu=\alpha_{\rm vis} h^2 \Omega$  \citep{Shakura1973}. A moderate $\alpha_{\rm vis}=0.01$ is adopted to ensure that both planets do not open deep gaps in the disc.

%\begin{figure}
%\centering
%\includegraphics[clip=true, width=0.45\textwidth]{res_ang.pdf}
%\caption{The resonance angle evolution for the inner object and the outer one. The inner object shows a finite resonance angle around 5000 orbits while the outer object still circulates around the full orbital phase (The resonant angle for the outer resonance is defined as $Q =2\lambda_A-\lambda_B-\varpi_A$). The two vertical dotted lines show 1000 and 5000 orbits just before and after the 2:1 MMR, same as Fig.~\ref{fig:orbit_sim}. Note that the times in this figure are all measured in unit of the orbital period at $r=r_{0}$.} 
%\label{fig:res_ang_sim}
%\end{figure}

A pair of planets is orbiting around the central star at a fixed orbit with the orbital semi-major axis $a_{\rm B}=1.0\ r_{0}$ and $a_{\rm A}=2^{2/3}\left( \frac{1+\mu_{\rm B}}{1+\mu_{\rm A}}\right)^{-1/3}r_{0}$,  where $\mu_{\rm B}=10^{-5}$ and $\mu_{\rm A}=3\times10^{-5}$ are the mass ratio of the inner and outer planets. This orbital configuration ensures that the two planets are in 2:1 MMR. Other planet pairs with 3:2 MMR is  discussed in Sec.~\ref{sec:32res} with $a_{\rm A}$  modified accordingly. Such small mass ratios for both planets are chosen to ensure that the tidal response of the disc is in the Type I regime. Note that the theory in the Sec.~\ref{sec:q} should also be valid even though the mass ratio of the pair does not satisfy $\mu_{\rm A}\gg \mu_{\rm B}$.
For our fiducial case, the inner planet is fixed at an eccentricity of $e_{\rm B}=0.03$, while the outer one moves along a nearly circular orbit with $e_{\rm A}=0.001$. Both planets have a longitude of pericenter of $\varpi_{\rm A}=\varpi_{\rm B}=0.0$. The dependencies of $\varpi_{\rm B}$ and $e_{\rm B}$ have also been explored.
The back reaction of disc force and mutual interaction for each planet is turned off such that both  planets are fixed at the prescribed orbits described above. In this way, we can evolve both planets in a controlled manner, and make better comparison with our theory.
With this kind of setup, both planets have a resonance angle of $Q=0.0$.

The gravitational potential for each planet of the pair is described as Eq.~\eqref{eq:potential}.
%
%\begin{eqnarray}
%    \Phi_{s} =   -\frac{G \mu_{s}M}{(\left|\bm{r}_{{\rm p},s}-\bm{r}\right|^2+\epsilon^2)^{1/2}}
%    + \mu_{s} \Omega_{{\rm p},s}^{2} \bm{r}_{{\rm p},s} \cdot \bm{r}
%    ,
%    \label{eq:potential}
%\end{eqnarray}
%where $s=A,B$ for each planet, the first term is the direct potential from the own planet, and the second term is the indirect potential arising from our choice of coordinate system.
We apply a gravitational softening, with length scale $\epsilon=0.6h$, to each planet's potential. 

We evolve the disc in a 2D cylindrical coordinate with a uniform radial grid of $n_{r}=384$ points in a radial domain between $[0.2,4.0] r_0$, and a uniform azimuthal grid with $n_{\phi}=640$ points. A convergence test with higher resolutions $[n_{r},n_{\phi}]=[512,1024]$ does not show significant impact on the dynamics of the planet pair. 
Two wave-killing zones are applied at the both the inner and outer radial edge to remove  wave reflections near both boundaries \citep{deValborro2006}.

\begin{figure*}
\centering
\includegraphics[clip=true, width=0.45\textwidth]{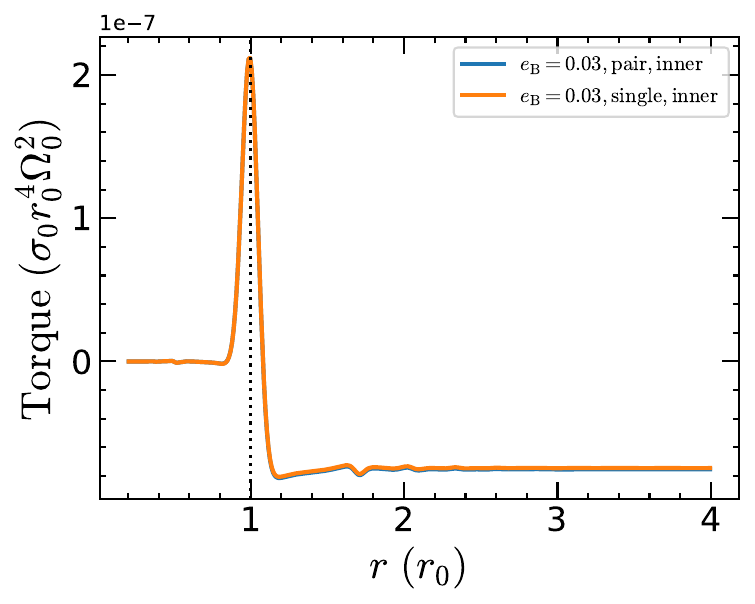}
\includegraphics[clip=true, width=0.45\textwidth]{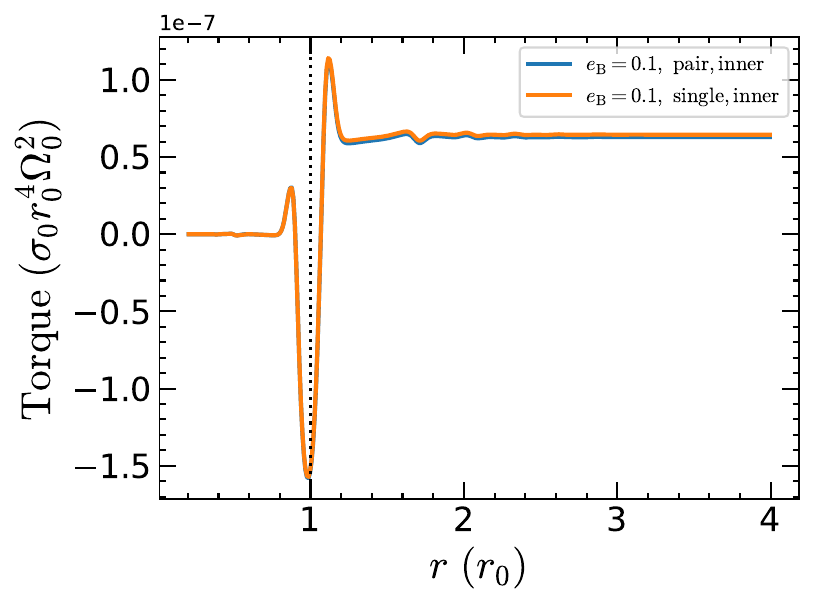}
\includegraphics[clip=true, width=0.45\textwidth]{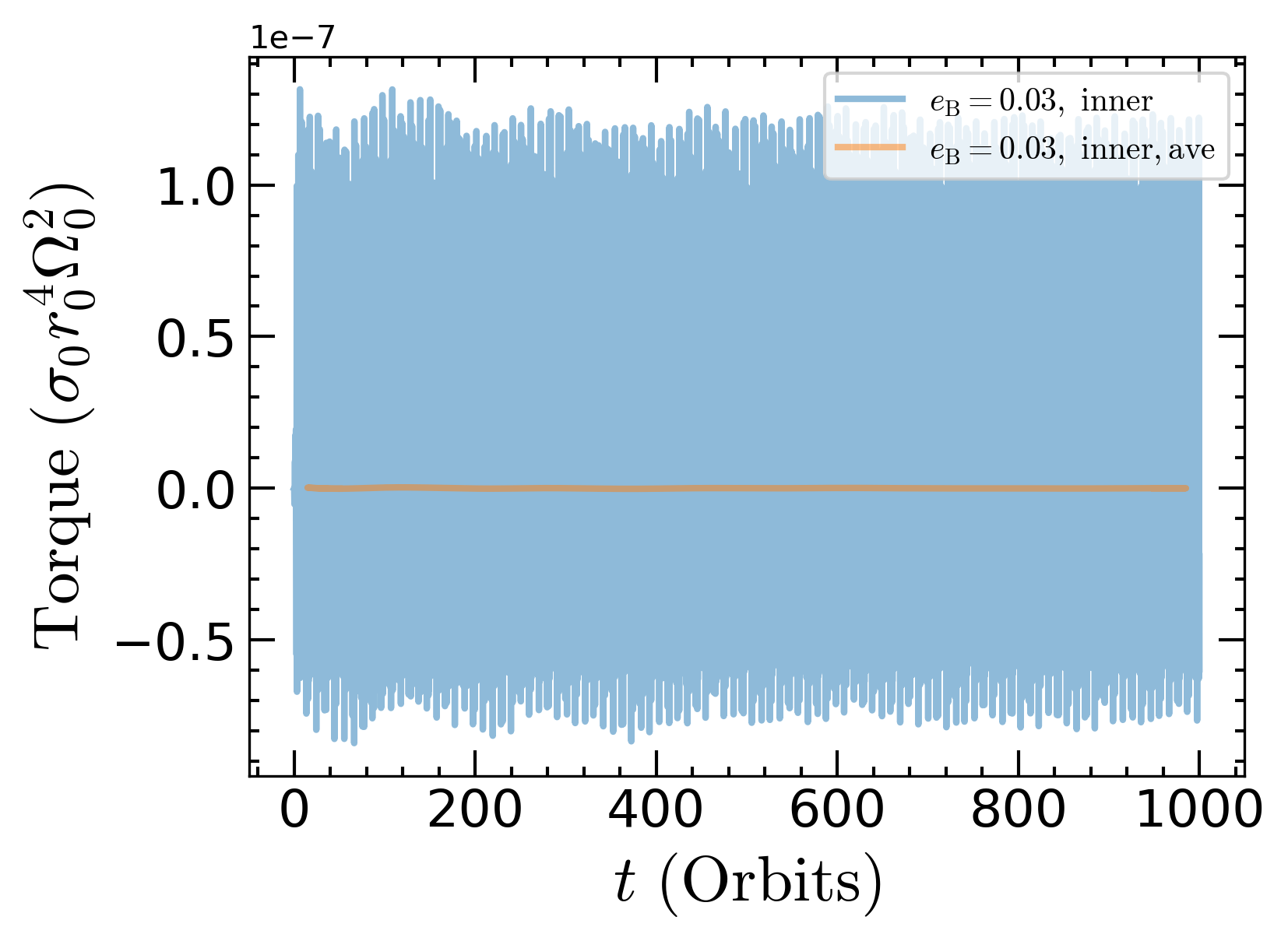}
\includegraphics[clip=true, width=0.45\textwidth]{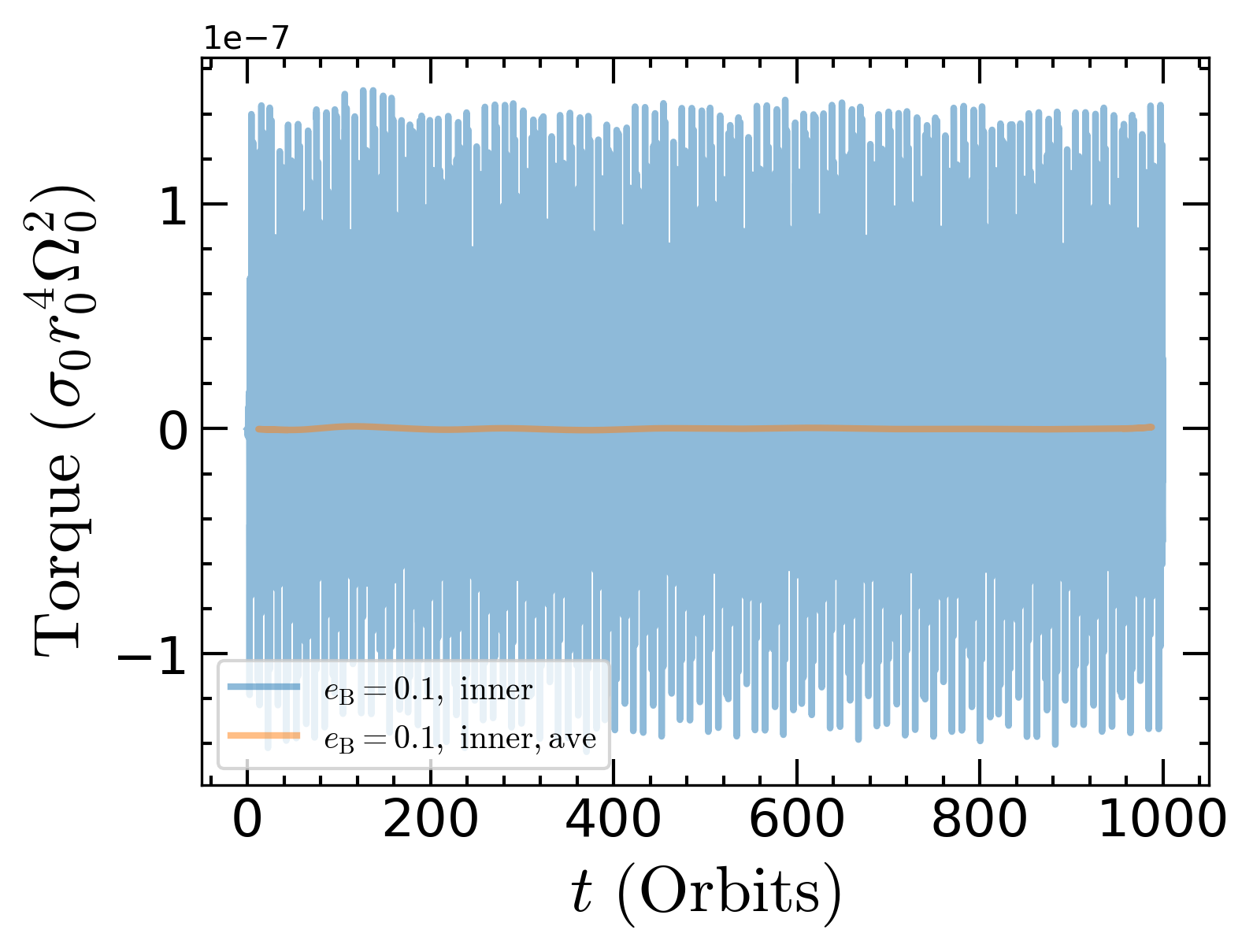}
\caption{Upper Left: the cumulative gravitational torque on the embedded objects as a function of $r$ for different models with the inner planet eccentricity $e_{\rm B}=0.03$. The blue line shows the torques for the inner planet based on the planet pair simulation, while the orange line shows the torque from the inner single planet case. 
The dotted line indicates the location of the inner planet.
The cumulative torque is obtained by integrating the differential torque density over $r$ from the inner boundary to a given $r$.
%so the total gravitational torque for each embedded object is the value at the large enough $r$ (e.g., $r>3r_{0}$).  
All the torque calculations are time-averaged within 100 orbits around 1000 orbits 
Lower Left: the time evolution of the interference torque on the inner planet location. The blue line shows the instantaneous torque while the orange one is the time-averaged value. 
%We can see that the time-averaged value is well consistent with zero although the instantaneous torque is strong. 
We can see that the single planet simulation can well reproduce the torques in the planet pair simulations, which means no net interference torque on the inner planet averaged over time when the pair is out of MMR.
Right panels: similar to the left ones except that the inner planet eccentricity is fixed at $e_{\rm B}=0.1$.
} 
	\label{fig:torque_noMMR}
\end{figure*}

\begin{figure}
\centering
\includegraphics[clip=true, width=0.45\textwidth]{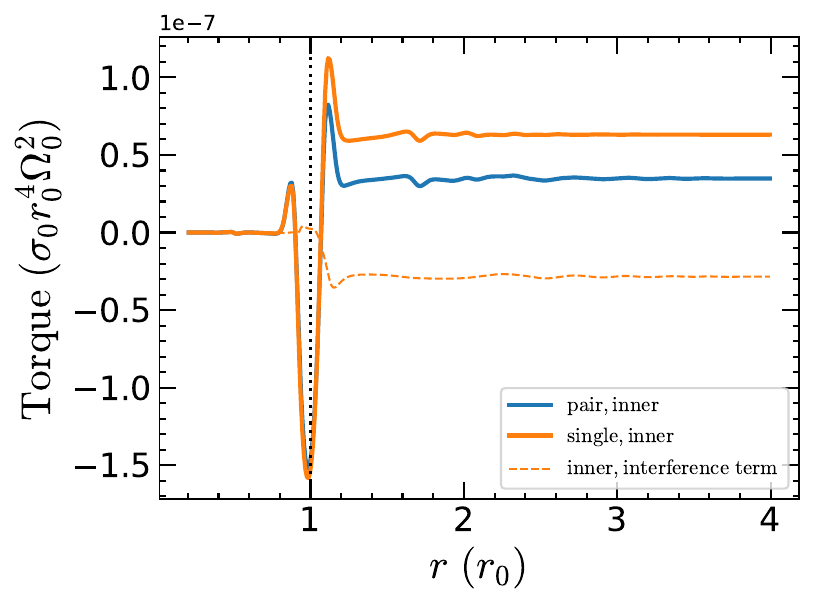}
\caption{Similar to the upper right panel of Fig.~\ref{fig:torque_noMMR} with $e_{\rm B}=0.1$, but for the case when the planet pair is in 2:1 MMR with a resonance angle of $Q=0$. The dashed line shows the time-averaged interference torque due to the density wave generated by the outer planet acting on the inner planet.
All the torques are averaged within 100 orbits around 1000 orbits. 
We can see that after the planet pair being captured into MMR, there exists significant torque difference for the inner object between the single planet and planet pair simulations. } 
	\label{fig:torque_MMR}
\end{figure}

\subsection{Simulations Results}

\subsubsection{Non-Resonant Pairs}

Before we probe the existence of interfering density waves,  we first run one planet pair simulations with their orbital period ratio is away from any MMR, e.g., we set $a_{\rm B}=1.0$, and $a_{\rm A}=1.7$.
Under such circumstance, the resonance angle of  both planets circulates about the full $2\pi$, regardless of the value of $\varpi_{\rm A}$ and $\varpi_{\rm B}$. The time-averaged interference torque should vanish due to the $\cos Q$-like dependence.
To verify this, we further run one single planet simulations with its orbit fixed in the same radius as the inner planet of the pair case. 

We can then track the torque acting on the inner planet for both cases. The time-averaged radial profile of the torque is shown in the upper left panel of Fig.~\ref{fig:torque_noMMR}.
We can see that the single planet can well reproduce the torque acting on the inner planet for the pair case, which means that the time-averaged torque induced by the density perturbation of the the outer planet vanishes as expected. To further confirm this, we also run another single planet simulation with its orbit being same as the outer planet of the planet pair, and a zero-mass inner passive particle is added to follow in the inner planet orbit. The torque acting on the passive inner planet, which is due to the density perturbation induced by the outer planet, can be instantaneously tracked. The time evolution and time-averaged value of the torques are shown in the lower left panel of Fig.~\ref{fig:torque_noMMR}. It can be clearly seen that the time-averaged interference torque due to the outer planet is consistently zero although the instantaneous torque magnitude is large. This essentially demonstrate the net zero interference torque for the non-MMR pair, as expected from the $\cos Q$-like dependence.

Furthermore, we test the dependence on the inner planet eccentricity, as shown in the right panels of Fig.~\ref{fig:torque_noMMR}. Here we increase the inner planet eccentricity to $e_{\rm B}=0.1$. Similar to the low eccentricity case in the left panels, the gravitational torques from the single planet located in the inner orbit are consistent with those of the planet pair case after performing the  time-averaged procedure. Note that the cumulative torque on the planet is positive. This is because the gravitational torque significantly depends on the eccentricity. For low eccentricity orbits, e.g. the  $e_{\rm B}=0.03$ case, the cumulative torque is indeed negative. When the eccentricity is much higher, e.g. the $e_{\rm B}=0.1$ case, the cumulative torque becomes positive which mainly damps out the eccentricity, with the semi-major axis still {\it decreasing} in time.
%one for the orbital migration and the other
%for  the eccentricity damping when the orbital eccentricity $e_{\rm B}$ of the inner planet is considerably large, with most of the positive torque being contributed to the damping of the orbital eccentricity. The contribution to the eccentricity evolution is negligible for a much smaller eccentricity of $e_{\rm B}=0.03$. 
%As a result, the cumulative torque for $e_{\rm B}=0.03$ is negative, while it is positive for $e_{\rm B}=0.1$.}

\subsubsection{Planet Pairs with 2:1 MMR}

Now we explore the interference term when the planet pair is fixed at the 2:1 MMR configuration and a resonance angle for the inner planet at $Q=0.0$, as described above.
We carry out two single-planet simulations with each one set at the orbital configuration of the inner or outer planet of the pair, respectively.
We can then calculate the torques on the inner planets for both the single planet and planet pair simulations to probe the existence of interfering density waves. 
%as motivated by Eq.~\eqref{eq:Tbinnu}.
The cumulative radial profile of the torques for single and pair cases with $e_{\rm B}=0.1$ are shown in Fig.~\ref{fig:torque_MMR}.
There is a remarkable difference for the time-averaged torque on the inner planet between the planet pair and single planet cases. This torque difference closely matches the interference torque due to the outer planet's perturbation, shown as the dashed line in the same plot. This time-averaged interference torque  is not vanishing due to the approximately constant $Q$ in the MMR regime.

\begin{figure}
\centering
\includegraphics[clip=true, width=0.45\textwidth]{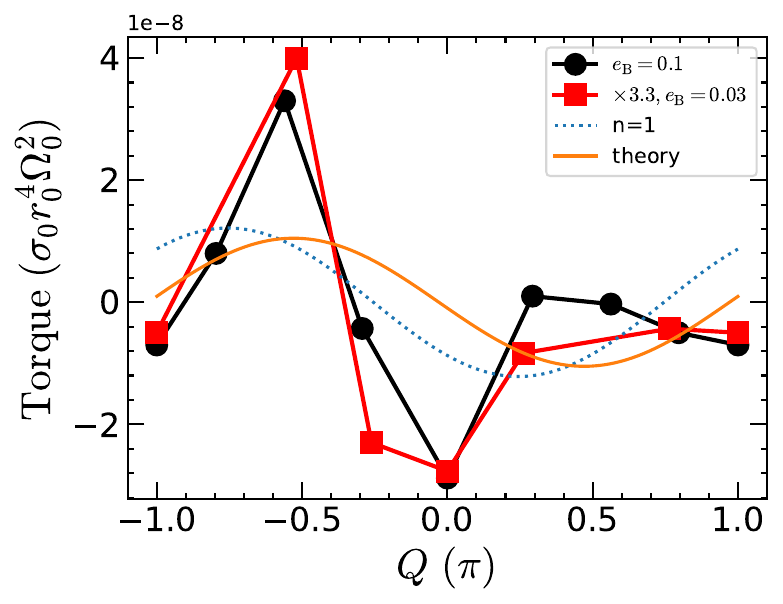}
\caption{The interference torque as a function of the resonance angle for the inner planet. The black line shows the case where the inner planet has an eccentricity $e_{\rm B}=0.1$ while the red line corresponds to the case where $e_{\rm B}=0.03$. Considering the linear dependence of the interference torque on $e_{\rm B}$, we scale up the red line by a factor of $3.3$. We can see that this linear dependence of the interference torque on $e_{\rm B}$ is quite well. 
The dotted line is the projection onto the $n=1$ component for the simulated interference torque, and the solid orange line is the theoretical predication to compare with the dotted line.
} \label{fig:torque_Q}
\end{figure}

The difference of the torque between the planet pair and single planet cases shown in Fig.~\ref{fig:torque_MMR} can then be used to compare with the analytical theory for the additional torque associated with the interfering density wave. 
The torque due to the interfering density wave for the inner planet measured directly from simulation (the difference between the blue and orange orange lines in Fig.~\ref{fig:torque_MMR}) is about $\sim-2.9\times10^{-8}$ (in code unit), also shown as the black line in Fig.~\ref{fig:torque_Q} with $Q=0.0$.
The location of the torque jump is located exactly at the inner Lindblad resonance for the outer planet of the pair.
In order to explore the dependence of the resonance angle $Q$, we further run a few simulations with different $\varpi_{\rm A}$ and $\varpi_{\rm B}$. The results of the interference torque for different $Q$ with $e_{\rm B}=0.1$ are shown as the black line in Fig.~\ref{fig:torque_Q}. 
We can see that the sinusoidal-like pattern of the interference torque with respect to $Q$ can be reproduced from our numerical simulations, although the pattern does not exactly follow the sinusoidal dependence with a single frequency in $Q$. This may suggest that there are high orders of harmonics of $Q$ in the interference torque, which are neglected in the analytical theory which only considered $|\ell-m| \le 1$ case.
To filter out the high-order harmonics, we make the projection for the simulation data onto the $n=1$ component of $A_{\rm Tb}\sin nQ + B_{\rm Tb}\cos nQ$. The coefficients $A_{\rm Tb}$,$B_{\rm Tb}$ are obtained from the integration $\frac{1}{\pi}\int_{0}^{2\pi}T_{\rm B,sim}(Q)\sin Q dQ$ and $\frac{1}{\pi}\int_{0}^{2\pi}T_{\rm B,sim}(Q)\cos Q dQ$ respectively. The $n=1$ component of the interference torque from the simulation data is indicated as the blue dotted line in Fig.~\ref{fig:torque_Q}.

In the limit of the disc self-gravity is negligible, our analytical theory in Sec.~\ref{sec:alt} predicates that the additional $n=1$ component torque from the interfering density wave can be estimated based on Eq.~\eqref{eq:tb2} assuming a Keplerian disc, which is shown as the orange line in Fig.~\ref{fig:torque_Q}.
This is roughly consistent with that expected from our simulation results both in magnitude and phase dependence, although some discrepancy exists in the phase. There are still important differences in the assumption made in the analytical theory and the numerical simulation shown here, despite both are using the same smoothing scheme. First, the theory assumes polytropic equation of state, while  the disc in our simulations is locally isothermal. Second, the theory assumes no viscosity in the disc, but a non-negligible $\alpha$-viscosity is adopted in our simulations to maintain the Type I migration regime.
The slightly non-Keplerian disc due to the pressure gradient will shift the resonance location from exactly 2:1 MMR location, which may impose a considerable effect on the torque.

We have confirmed from simulations that this interfering density wave is weaker for the outer planet.

\begin{figure}
\centering
\includegraphics[clip=true, width=0.45\textwidth]{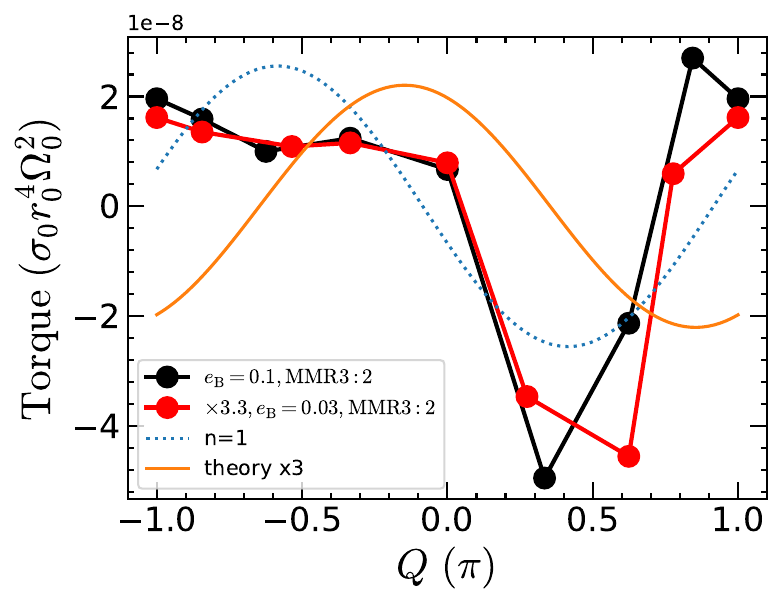}
\caption{Similar to Fig.~\ref{fig:torque_Q} but for 3:2 MMR. Note that the orange curve has been multiplied by a factor of 3 to match the magnitude of $n=1$ curve from the simulation data.} 
	\label{fig:torque_Q32}
\end{figure}

\begin{figure}
\centering
\includegraphics[clip=true, width=0.45\textwidth]{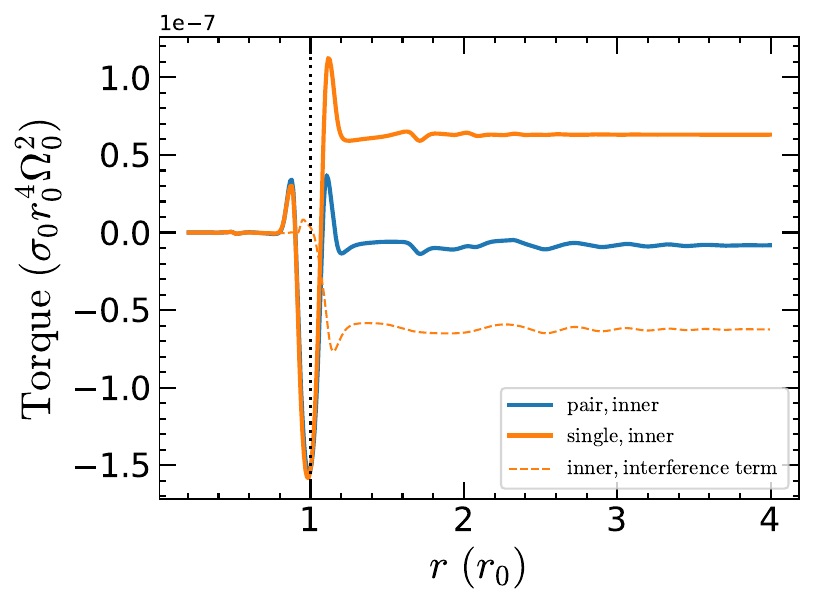}
\caption{Similar to Fig.~\ref{fig:torque_MMR}, but with the outer planet mass ratio of $\mu_{\rm A}=6\times10^{-5}$, which is a factor of two larger. } 
\label{fig:torque_MMR_m2}
\end{figure}
 
%In addition, another term of $-1.04 \left( \frac{h}{a_B}\right )^{1/3}\cos(Q-0.52)$ is neglected in Eq.~\eqref{eq:Tbinnu}, which will bring the theoretical value closer to our simulation results.

\subsubsection{Other Parameter Dependence}\label{sec:32res}

It is interesting to further explore parameter dependence of the interface torques. 
The first one is the linear dependence on the inner planet $e_{\rm B}$ which appeared in $\Psi_{B,l,m}$ in Equation~\ref{eq:tb2}.
To this end, we carry out a few simulations with $e_{\rm B}=0.03$, and the results for the interface torques are shown as red dots in Fig.~\ref{fig:torque_Q}. 
On the one hand, as we scale the overall torque amplitude for the simulations with two different eccentricities, we find that the interference torque is approximately  linear in $e_{\rm B}$, which is consistent with the expectation from the linear theory.
On the other hand,
the relative contribution from high orders of harmonics seems to be weaker for the lower eccentricity case as we compare the simulations with two separate eccentricities.

Second, we also perform several simulations for planets locked in 3:2 MMR for both $e_{\rm B}=0.1$ and $e_{\rm B}=0.03$ with different $Q$. The results are shown in Fig.~\ref{fig:torque_Q32}. The $n=1$ component of the simulated interference torques are shown as the dotted line, and the theoretical predication based on Eq.~\eqref{eq:tb2} is represented as the solid orange line. The linear dependence on $e_{\rm B}$ is quite similar to the 2:1 MMR case. However, in order to match the torque magnitude for the simulation data, the theoretical curve in Fig.~\ref{fig:torque_Q32} has been multiplied by a factor of 3. The phase offset between the theory and simulation curves is also larger than that of 2:1 MMR. This suggests that there may exist  important factors that lead to such deviation, with some possible reasons  discussed above.

Third, we further examine  the mass ratio dependence for the interference torque, which should be proportional to the mass of the both planet (ref. to Eq.~\eqref{eq:tb2} or Eq.~\eqref{eq:tb2_re} for the dependence on $\Psi_{A,m,m}$ and $\Psi_{B,l,m}$). We increase the mass of the outer planet by a factor of two, and fixed the other model parameters as in Fig.~\ref{fig:torque_MMR}. The results are shown in Fig.~\ref{fig:torque_MMR_m2}. We can see that the interference torque is quite similar to that of Fig.~\ref{fig:torque_MMR} except that the magnitude is also a factor of two larger, as expected.

%To probe the dependence on different MMR pairs, we perform simulations with a few  different eccentricity $e_{\rm B}$ for the inner planet and different $\varpi_{\rm B}$ which results in different resonance angles for the inner planet. The results are shown in Fig.~\ref{fig:torque_Q}. 
%We can see that the sinusoidal-like pattern of the interference torque with respect to $Q$ can be reproduced from our numerical simulations, although the pattern does not exactly follow the sinusoidal dependence with a single frequency in $Q$. This may suggest that there are high orders of harmonics of $Q$ in the interference torque, which are neglected in the analytical theory which only considered $|\ell-m| \le 1$ case. 

%The relative contribution from high orders of harmonics seems to be weaker for the lower eccentricity case as we compare the simulations with two separate eccentricities. On the other hand, as we scale the overall torque amplitude for the simulations with two different eccentricities, we find that the interference torque is approximately  linear in $e_{\rm B}$, which is consistent with the expectation from the linear theory (c.f. Eq.~\eqref{eq:psib_soft}). %This is demonstrated by the linear scaling of red and black lines in Fig.~\ref{fig:torque_Q}, where the linear dependence is well reproduced by our simulations.

In summary, by considering  the comparison between the numerically extracted torque and analytically predicted torque, i.e.,  the location of the torque jump which locates at the inner Lindblad resonance, the torque amplitudes, the linear dependence on $e_{\rm B}$, planet mass, and the sinusoidal-like pattern of the additional torque term,
it is fair to state that the existence of the interference torque is evident.

%One possible origin of this discrepancies may arise from the small eccentricity approximation assumed in obtaining $\tilde{\Psi}_{B,m-1,m}$ and Eq.~\eqref{eq:Tbinnu}. The backreaction of planet B to the local disc profile seems to be small in the current set-up.

%The analytical theory in the previous section uses ``ad hoc" vertical average prescription following \citetalias{ward1988disk}, so that the apparent divergence in $\Psi_{B,m-1,m}$ in the two-dimensional theory can be regularize. If we were to apply $\Psi_{B,m-1,m}$ in Eq.~\eqref{eq:bin3} we would obtain a positive interference torque. The fact these two different smoothing schemes in the two-dimensional theory giving quite different predictions in the backreaction torque means that the actual physical torque may sensitively depend on the local three-dimensional disc structure. A more accurate treatment should be solving the three-dimensional model and considering the vertical modes similar to the analysis in \cite{tanaka2004three}. This calculation is beyond the scope of this work and needs detailed studies in the future. 

\section{Modified  Resonant Dynamics}\label{sec:mod}

\begin{figure}
	\centering
    \includegraphics[clip=true, width=0.45\textwidth]{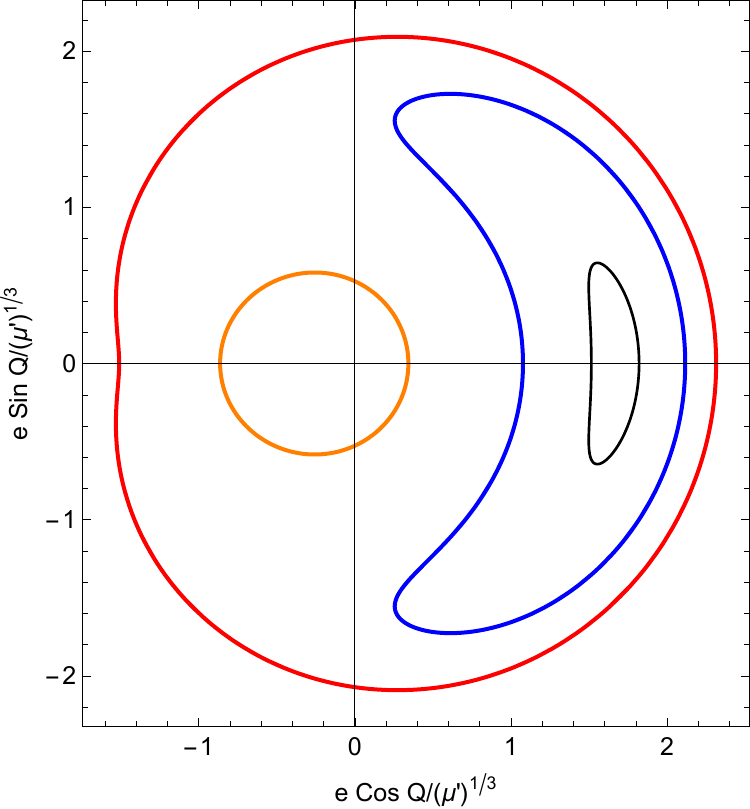}
	\caption{The phase space trajectories of a $2:1$ resonance with $k =2 k_{\rm crit}$.} 
	\label{fig:phase1}
\end{figure}

%\begin{figure}

%            \includegraphics[width=0.4\textwidth, height=0.3\textwidth]{eccentricity.pdf}
%               \label{fig:eccentricity}

%            \includegraphics[width=0.4\textwidth, height=0.3\textwidth]{phi.pdf}
%             \label{fig:phi}
        
%    \caption{
%     The evolution of $e$ and $Q$ for the late-time behavior of the type-$(a)$ orbit, which is well described by the attractor solution in Eq.~\eqref{eq:att1} .
%     }
%    \label{fig:ephi}
%\end{figure}

As discussed in Sec.~\ref{sec:q}, for a pair of massive objects embedded in an accretion disc and trapped within an MMR, the interfering density waves produces a backreaction torque that depends on the resonant angle $Q$. Such $Q$-dependence likely introduces qualitatively different resonant dynamics from those systems involving only gravitational interactions, or those only include constant (in the resonant angle) migration torques. We address the associated dynamical signatures in this section.

Without including the torque from interfering density waves, the equations of motion for the pair of objects considered in Sec. \ref{sec:q} has already been discussed in \citetalias{Goldreich:2013rma}, in which case the outer object A stays at a fixed circular orbit ($\dot{r}_A=0$) and the inner object B migrates outwards. The equations of motion including the effect of interfering density waves for $n=\dot{\lambda}_B$, $e$ (the eccentricity of the inner planet $e_{\rm B}$) \footnote{We drop out the subscript ``B" hereafter unless otherwise noted since we only focus on the dynamics of the inner planet.}  are 
\begin{align}\label{eq:dne}
\dot{n} & = 3 (m-1) \beta_0 \mu' e n^2 \sin Q -\frac{n}{\tau_n}+p \frac{e^2 n}{\tau_e}- \frac{6e n\cos Q}{\tau_0}  \nonumber \\
\dot{e}& = \beta_0 \mu' n \sin Q -\frac{e}{\tau_e}-\frac{\cos Q}{\tau_0}
\end{align}
where $\beta_0$ is approximately $0.8 (m-1)$, $\mu' :=M_A/M, \mu:=M_B/M$,  the $p$-related term may be contributed by remote first-order Lindblad resonances and corotation resonances. For illustration of principles we take the same value $p=3$ as used in  \citetalias{Goldreich:2013rma}. On the other hand, the change rate of mean motion and eccentricity due to single-body migration torques approximately scale as
\begin{align}\label{eq:tauscaling}
\frac{1}{\tau_n} \sim \mu \frac{\sigma a^2}{M} \left ( \frac{a}{h}\right )^2 n, \quad \frac{1}{\tau_e} \sim \mu \frac{\sigma a^2}{M} \left ( \frac{a}{h}\right )^4 n\,.
\end{align}

The resonant dynamics can be determined by combining Eq.~\eqref{eq:dne}  with the definition that $Q =m \lambda_A -(m-1) \lambda_B-\varpi_B$ and the equation of motion for $\varpi_B$:
\begin{align}
\dot{\varpi}_B = -\frac{\beta_0 \mu'}{e} n \cos Q\,.
\end{align}

\subsection{Without constant migration torques}\label{sec:nomig}

\begin{figure*}
    \centering
  \subfigure[]{       
            \includegraphics[width=0.4\textwidth, height=0.3\textwidth]{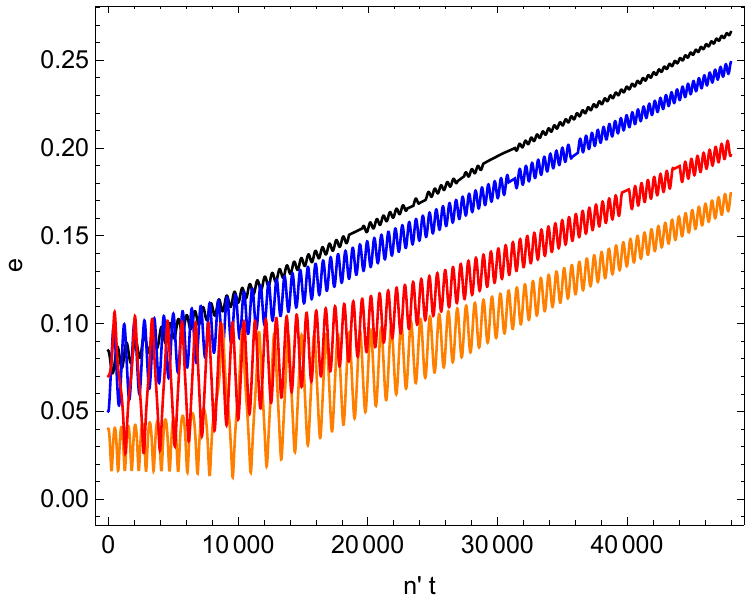}
               \label{fig:phase2}
        }~\subfigure[ ]{       
            \includegraphics[width=0.4\textwidth, height=0.3\textwidth]{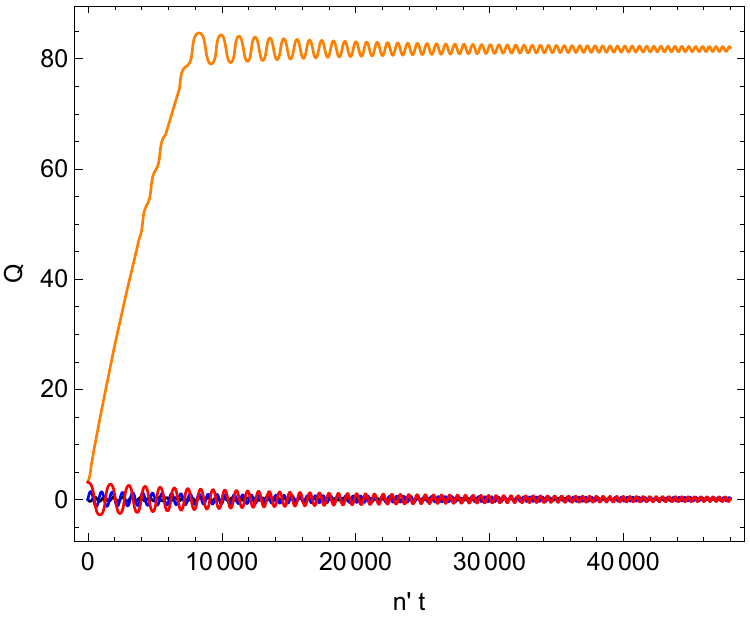}
             \label{fig:phase3}
        }
         \subfigure[]{       
            \includegraphics[width=0.4\textwidth, height=0.3\textwidth]{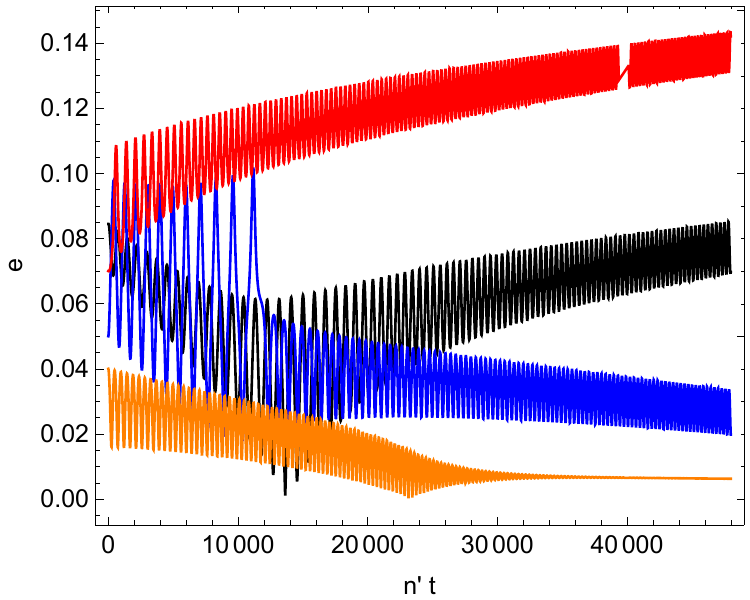}
               \label{fig:ecombine}
        }~\subfigure[ ]{       
            \includegraphics[width=0.4\textwidth, height=0.3\textwidth]{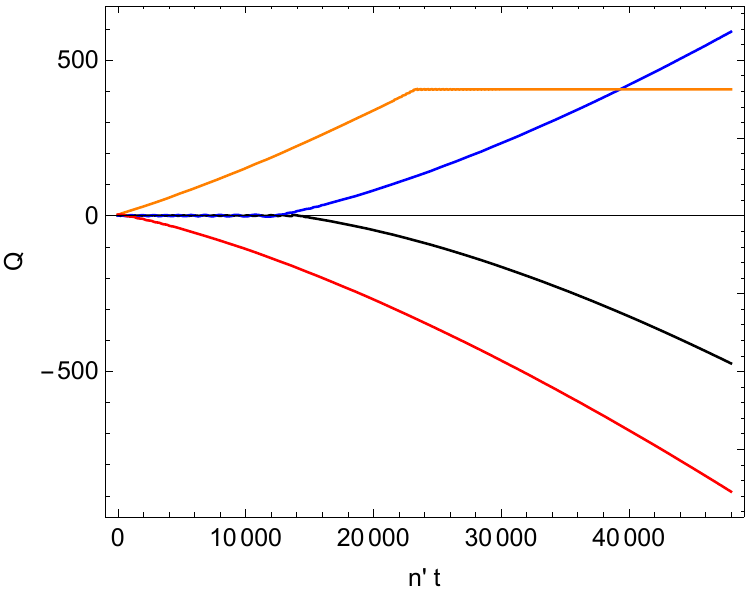}
             \label{fig:qcombine}
        }
    \caption{
     The evolution of the four trajectories presented in Fig.~\ref{fig:phase1} with $\tau_0 \beta_0 \mu' n' =20$ (top panels) and $\tau_0 \beta_0 \mu' n' =-20$ (bottom panels). The color scheme is chosen such that it is consistent with the phase-space trajectories in Fig.~\ref{fig:phase1} as initial data.
     }
    \label{fig:array}
\end{figure*}

Let us first consider the case with the normal migration torques turned off, i.e. removing $\tau_e, \tau_n$-related terms in Eq.~\eqref{eq:dne}. Although this assumption is made to present a simplified  discussion on the resonant dynamics, it becomes a reasonable approximation if $\mu' \gg \mu$, although in this case $n'$ is generally time-dependent.
If the $\tau_0$-related term is also absent, the equations of motion is compatible with the resonant Hamiltonian  \citepalias{Goldreich:2013rma}:
\begin{align}
\mathcal{H} = k e^2-\frac{3}{4} (m-1)^2 e^4+ 2 \beta_0 \mu' e \cos Q
\end{align}
with
\begin{align}
k:= \frac{3}{2}(m-1)^2 e^2 -\frac{\beta_0 \mu'}{e} \cos Q+\frac{\dot{Q}}{n}
\end{align}
being a constant in time. The conjugate canonical variables are $Q$ and $e^2$. Defining a critical $k$ as $k_{\rm crit} = 3^{4/3} ((m-1) \beta_0 \mu')^{2/3}/2$, the phase space contains one stable fixed point for cases with $k < k_{\rm crit}$ and two fixed points plus one unstable fixed point for $k > k_{\rm crit}$ \citep{murray2000solar,Goldreich:2013rma}. In Fig.~\ref{fig:phase1} we show several representative phase space trajectories in terms of a set of re-scaled canonical variables $\{e \cos Q /\mu'^{1/3}, e \sin Q /\mu'^{1/3}\}$.
The blue and black curves are ``libration" trajectories  around a fixed point at $e_{\rm max} \approx 1.96 (\beta_0 \mu')^{1/3}$ on the positive side of the real axis.  The red curve is a large ``rotation" orbit. The orange curve is a ``rotation" trajectory around the other stable fixed point at $e_{\rm min} \approx 0.093 (\beta_0 \mu')^{1/3}$ on the negative side of the real axis.

Now with the $\tau_0$ term included, it contributes to another $Q$-dependent driving source that has $90$-degree offset from the mutual gravitational interaction. The ratio of magnitudes is
\begin{align}
\frac{1}{|\tau_0 \beta_0 \mu' n |} &\sim 0.4\left ( \frac{\sigma a^2/M}{10^{-3}}\right ) \left ( \frac{a/h}{10}\right )^2 \nonumber \\
& \sim  \left ( \frac{\alpha}{0.1}\right )^{-1} \left ( \frac{\dot{M}}{0.1 \dot{M}_{\rm Edd}}\right )^{-3} \left ( \frac{M}{10^5 M_{\odot}}\right ) \left ( \frac{a}{10^3 M}\right )^{5.5}
\end{align}
which may  be comparable to one depending on the properties of the disc and the location of the object, so that the resonant dynamics may be significantly affected. In the first line we have noticed that a significant fraction of protoplanetary discs in the Lupus complex observed by the Atacama Large Millimeter/Submillimeter Array has disc gas mass to star mass ratio around $10^{-3}$ \citep{ansdell2016alma}. In the second line we have assumed a $\alpha$ disc profile around a supermassive black hole \citep{Kocsis:2011dr}, with $\dot{M}$ being the accretion rate of the central black hole and $\dot{M}_{\rm Edd}$ being the Eddington accretion rate. 

In Fig. \ref{fig:array} we present the evolution in the regime that $|\mu' n \tau_0 |\gg 1$, so that the force due to interfering density waves is much weaker than the gravitational interactions between the two objects. In addition, as discussed in the Sec.~\ref{sec:nume}, the sign of the backreaction torque depends on the detail three-dimensional structure of the gas flow near the inner Lindblad resonance, which is not resolved in this study. As a result, we consider two cases with $\tau_0 \beta_0 \mu' n'=\pm 20$ respectively. One common feature of these two scenarios is that the numerical evolution no longer preserves the area in the phase space of the canonical variables, so the evolution is no longer Hamiltonian. 

In the case that $\tau_0 \beta_0 \mu' n' =20$, we find various initial data all lead to an asymptotic libration state with increasing eccentricity. In fact, on the right hand side of the $\dot{e}$ equation in Eq.~\eqref{eq:dne}, the time average of $\beta_0 \mu' n \sin Q$ is slight larger than the time average of $\cos Q/\tau_0$, so that the libration regime drifts towards to the right. On the other hand, in the case that $\tau_0 \beta_0 \mu' n' =-20$, we find different kinds of final states. For the black and red trajectories, the system evolve to a rotation state around the origin with increasing eccentricity in time (similar to the red curve in Fig.~\ref{fig:phase1} with increasing amplitude). The blue trajectory instead lands on another rotation state with shrinking eccentricity in time (similar to the orange curve in Fig.~\ref{fig:phase1} with decreasing amplitude).

Mathematically we can approximately describe the end state of the black and red trajectory as follows. Without the $\tau_0$-related terms, the rotational orbits around the origin can be written as
\begin{align}\label{eq:approx}
n &= n_0 +\epsilon \delta n \cos Q + \mathcal{O}(\epsilon^2), \nonumber \\
e &= e_0 +\epsilon \delta e \cos Q + \mathcal{O}(\epsilon^2),\nonumber \\
Q &=\omega_0 t +\mathcal{O}(\epsilon)
\end{align}
where $\epsilon$ is a book-keeping parameter and we only keep terms up to the linear order in $\epsilon$.  In addition, according to the fact that
\begin{align}
\dot{Q} = m n'-(m-1) n +\frac{\beta_0 \mu'}{e} \cos Q\,,
\end{align}
we can immediately identify that $\omega_0 = m n' -(m-1) n_0$. Together with Eq.~\eqref{eq:dne} (with $\tau_0,\tau_n,\tau_e$-related terms removed), we find that
\begin{align}\label{eq:deltane}
\delta n & = -\frac{3 (m-1) \beta_0 \mu'e_0 n^2_0}{\omega_0}, \nonumber \\
\delta e & = -\frac{\beta_0 \mu' n_0}{\omega_0}\,.
\end{align}

%These solutions fit reasonably well with late-time evolutions of the type $(b,c,d)$ trajectories. 
In order to describe the secular evolution, we can use the evolution of conserved quantities $k, \mathcal{H}$:
\begin{align}
\frac{d k}{d t} & = 3 e \dot{e}_{\tau_0} -\frac{2n'}{n^2} \dot{n}_{\tau_0} =\frac{3 e \cos Q}{\tau_0} \,, \nonumber \\
\frac{d \mathcal{H}}{d t} &=e^2 \frac{d k}{d t} +\frac{d e^2}{d t}\left ( k-\frac{3}{2}e^2+\beta_0 \mu'/e \cos Q\right ) \nonumber  \\
& = \frac{3 e^3 \cos Q}{\tau_0} -\frac{2 e \cos Q}{\tau_0} \left ( k-\frac{3}{2}e^2+\beta_0 \mu'/e \cos Q\right ) \nonumber \\
& =  -\frac{2 e \cos Q}{\tau_0} \left ( k-3e^2 \right ) -\frac{2 \beta_0 \mu' \cos^2 Q}{\tau_0} \,.
\end{align}

After plugging the approximate solution in Eq.~\eqref{eq:approx} and Eq.~\eqref{eq:deltane} and performing average over oscillation cycles in the resonant phase $Q$, we find that (with $m=2$)
\begin{align}
\langle \dot{k} \rangle = -\frac{3 \mu' \beta_0 n_0}{\omega_0 \tau_0} \left ( 1+\frac{3n_0 e^2_0}{2\omega_0}\right ), 
%\quad
%\langle \dot{\mathcal{H}} \rangle = 3e^2_0 \langle \dot{k} \rangle\,,
\end{align}
which is also useful for the analysis in Sec.~\ref{sec:ob}. 
%Only type $(a)$-like orbits maintain bound resonant angle with the influence of interfering density waves, which likely come from initial orbits with small eccentricities before the resonant capture. These orbits can still achieve resonant locking.
On the other hand, the orange curve eventually evolves to a state that 
%The figures suggest that there are two ``attractor solutions" for the new equation of motion. For example, the black curve (part $(a)$ of Fig. \ref{fig:array}) which originally circulates around the stable fixed point on the positve real axis now asymptotes to a point with decaying oscillations.
%Mathematically this attractor solution can be found by requiring that (see Fig.~\ref{fig:ephi})
\begin{align}\label{eq:att1}
 & \beta \mu' n \sin Q \approx \frac{\cos Q}{\tau_0}\,, \nonumber \\
 &m n'-(m-1) n +\frac{\beta \mu'}{e} n \cos Q \approx 0\,,
\end{align}
i.e., $\dot{e}, \dot{Q} \approx 0$, so that the quasi-stationary values of $e, Q$ can be determined as functions of $n$. Notice that here $n$ is still time dependent according to Eq.~\eqref{eq:dne} (with $1/\tau_e, 1/\tau_n$ set to be zero), so that this  point drifts in time. In addition, from the second line of Eq.~\eqref{eq:att1} one can find that at equilibrium the offset of period ratio $m n'/n-(m-1)$ is inversely proportional to the eccentricity, i.e., smaller eccentricity corresponds to larger offset of the period ratio (see also the related discussion in Sec.~\ref{sec:nume}). This is a rather general point as long as $Q$ is bounded.

%The blue curve (part $(b)$ of Fig. \ref{fig:array}) that initially circulates around the same fixed point with a larger amplitude, on the other hand, first migrates to the left similar to the black curve, but at some point switch to a new branch of orbits rotating around the origin with larger and larger amplitude, corresponding to a new ``attractor solution". This solution has growing $\mathcal{H}$ in time, which must be contributed by the interfering density wave term. In fact, we also find that the  orange and red  orbits (part $(c)$ and $(d)$ of Fig. \ref{fig:array}) directly evolve towards this attractor solution. The presence of this attractor solution shows that the MMR becomes more unstable with the interfering density waves because the extra term breaks the resonance locking for a fraction of parameters in the libration regime.

\begin{figure*}
    \centering
  \subfigure[]{       
            \includegraphics[width=0.4\textwidth, height=0.25\textwidth]{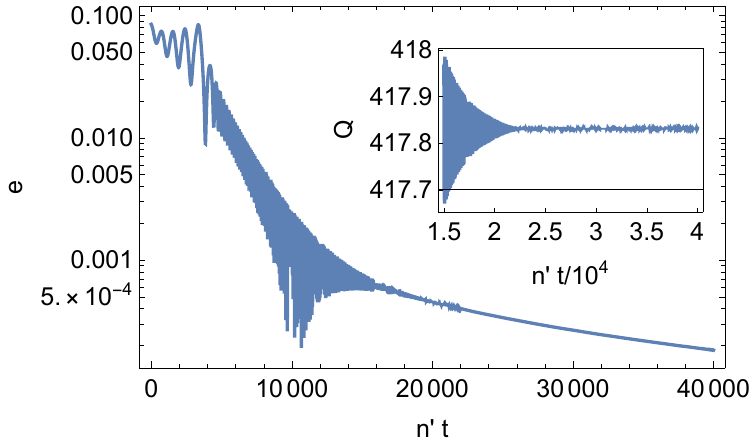}
              
        }~\subfigure[ ]{       
            \includegraphics[width=0.4\textwidth, height=0.25\textwidth]{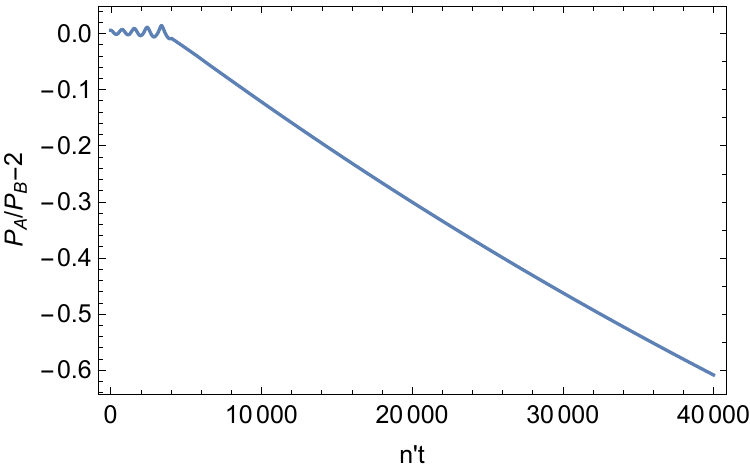}
             
        }
        \subfigure[]{       
            \includegraphics[width=0.4\textwidth, height=0.25\textwidth]{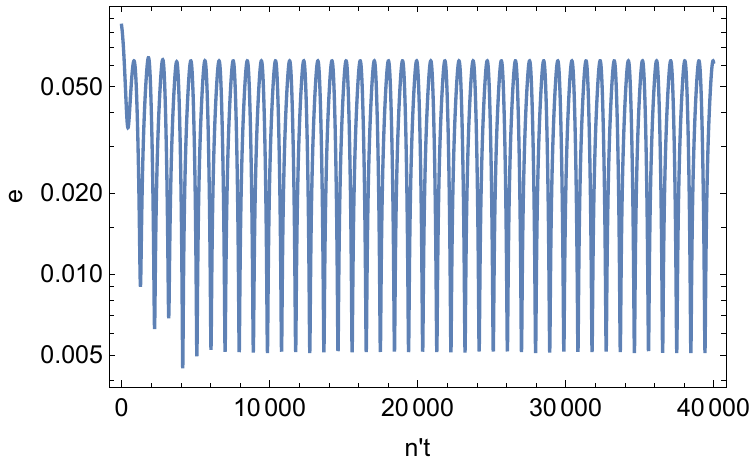}
               
        }~\subfigure[ ]{       
            \includegraphics[width=0.4\textwidth, height=0.25\textwidth]{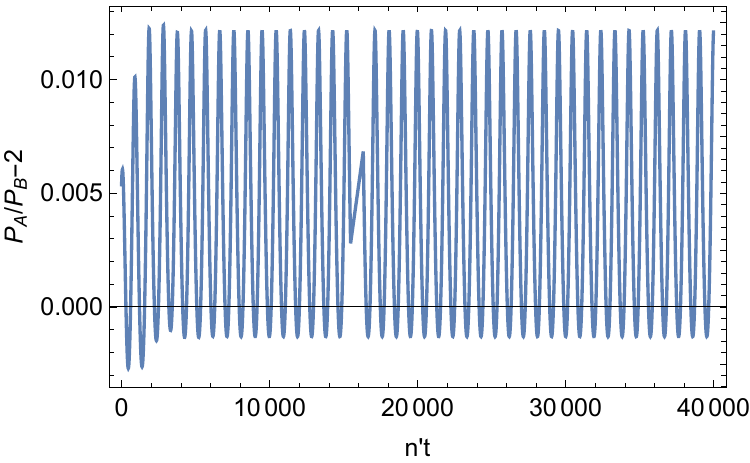}  
        }
         \subfigure[]{       
            \includegraphics[width=0.4\textwidth, height=0.25\textwidth]{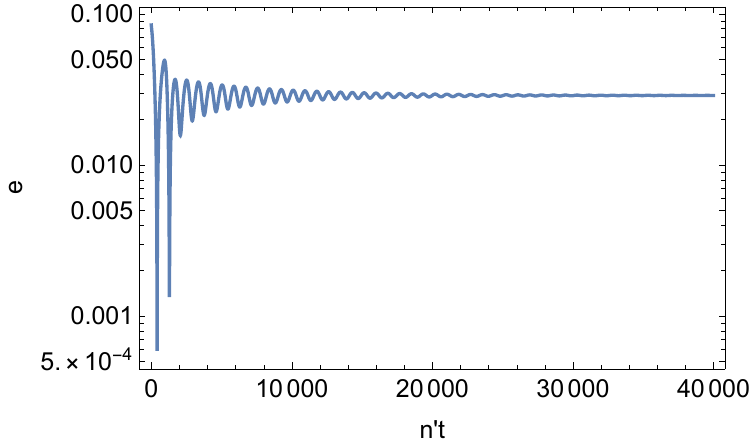}
               
        }~\subfigure[ ]{       
            \includegraphics[width=0.4\textwidth, height=0.25\textwidth]{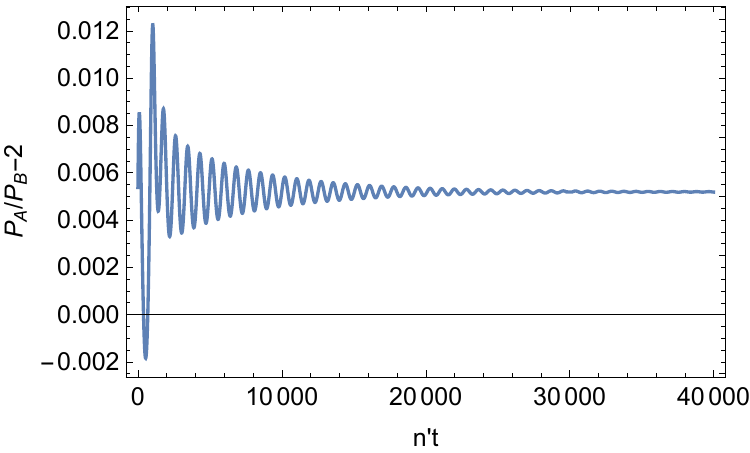}  
        }
    \caption{
     The evolution of eccentricities and periods  with $\tau_e = \tau_n/50$ (top row), $\tau_e =\tau_n/100$ (middle row) and $\tau_e= \tau_n/200$ (bottom row) respectively, according to Eq.~\eqref{eq:dne} but with $\tau_0$ terms turned off. The scaled mass of object A is assumed to be $\mu' =10^{-4}$ and $\tau_n$ is set to be $n' \tau_n =10^5$. The insect of panel (a) shows the evolution of the resonant angle $Q$, which is bounded at late times.
     }
    \label{fig:array2}
\end{figure*}

\begin{figure*}
    \centering
  \subfigure[]{       
            \includegraphics[width=0.4\textwidth, height=0.25\textwidth]{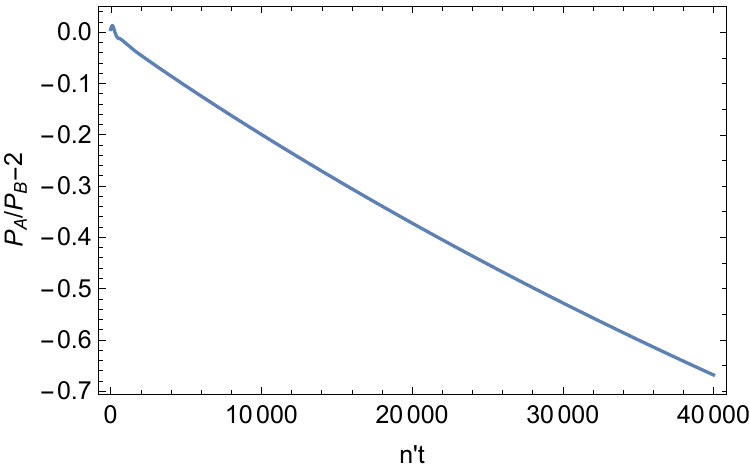}
              
        }~\subfigure[ ]{       
            \includegraphics[width=0.4\textwidth, height=0.25\textwidth]{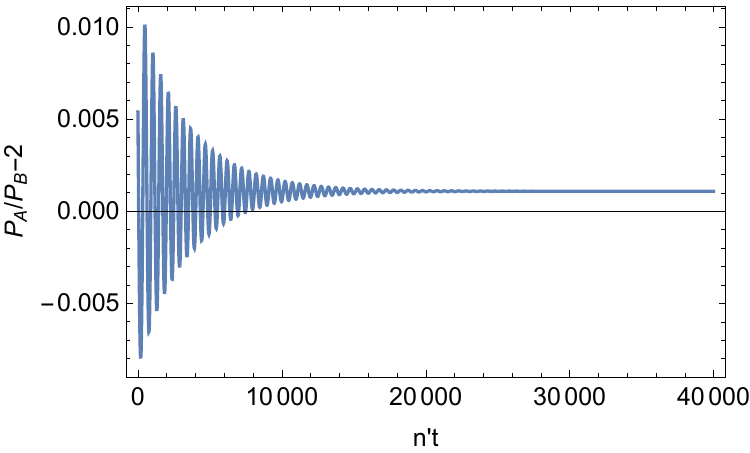}
             
        }
        \subfigure[]{       
            \includegraphics[width=0.4\textwidth, height=0.25\textwidth]{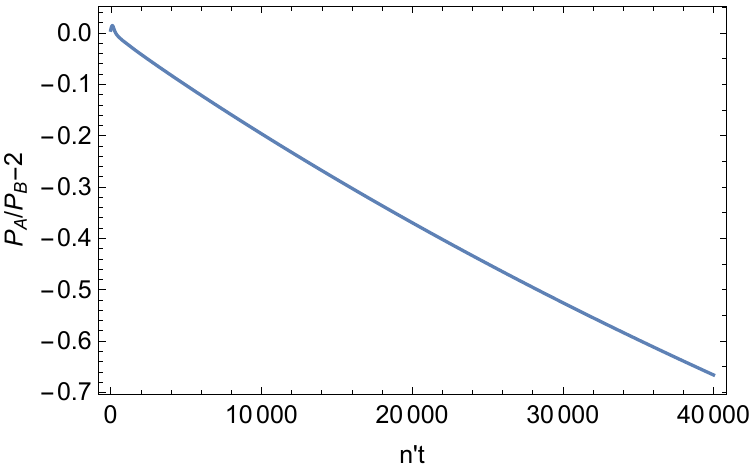}
               
        }~\subfigure[ ]{       
            \includegraphics[width=0.4\textwidth, height=0.25\textwidth]{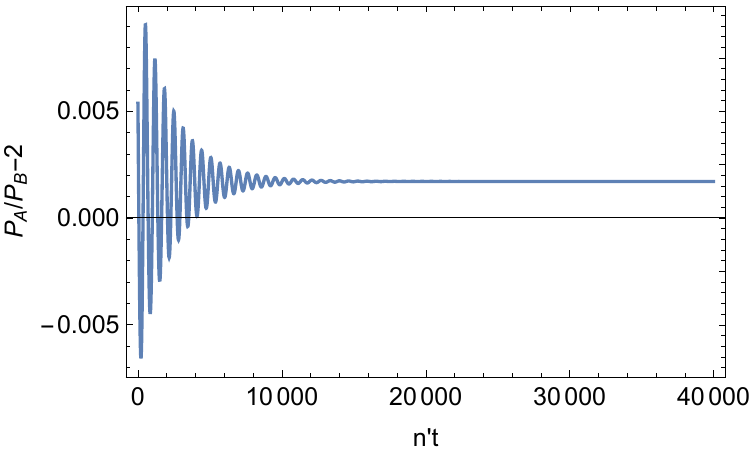}  
        }
         \subfigure[]{       
            \includegraphics[width=0.4\textwidth, height=0.25\textwidth]{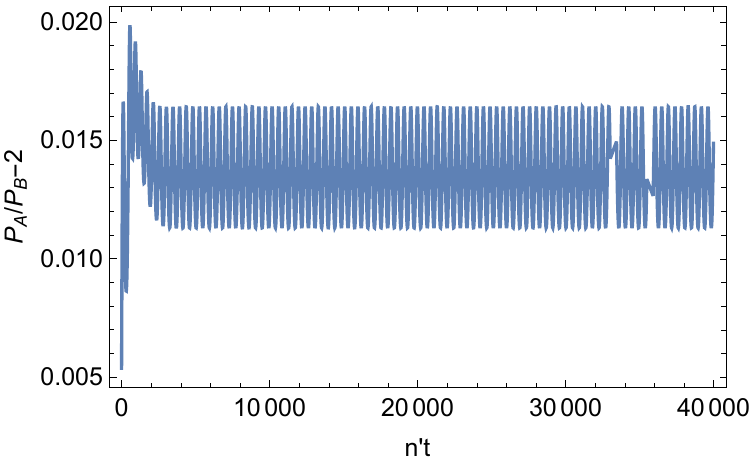}
               
        }~\subfigure[ ]{       
            \includegraphics[width=0.4\textwidth, height=0.25\textwidth]{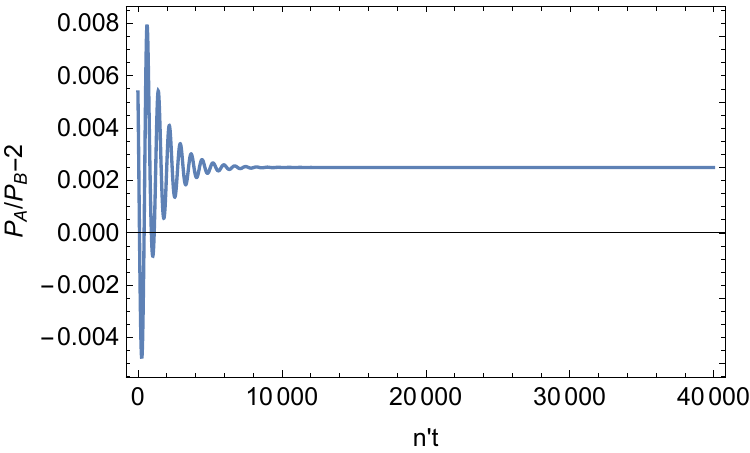}  
        }
    \caption{
     Similar set up as Fig.~\ref{fig:array2} but with $\tau_0$ set to be $-\tau_n/10$ on the left column and $\tau_n/10$ on the right column.
     }
    \label{fig:array3}
\end{figure*}

\subsection{With migration torques}\label{sec:3with}

\begin{figure*}
    \centering
  \subfigure[]{       
            \includegraphics[width=0.4\textwidth, height=0.25\textwidth]{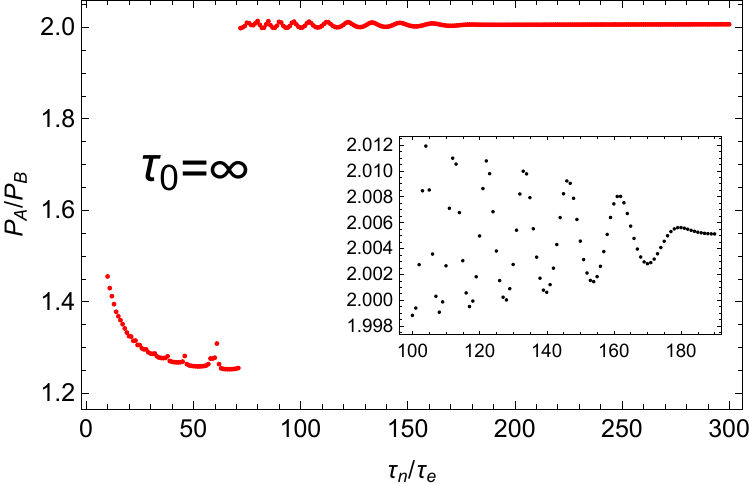}
              
        }~\subfigure[ ]{       
            \includegraphics[width=0.4\textwidth, height=0.25\textwidth]{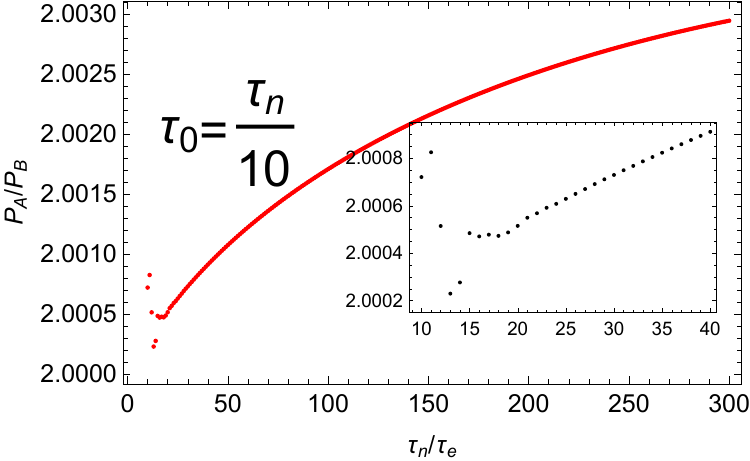}
             
        }
        \subfigure[]{       
            \includegraphics[width=0.4\textwidth, height=0.25\textwidth]{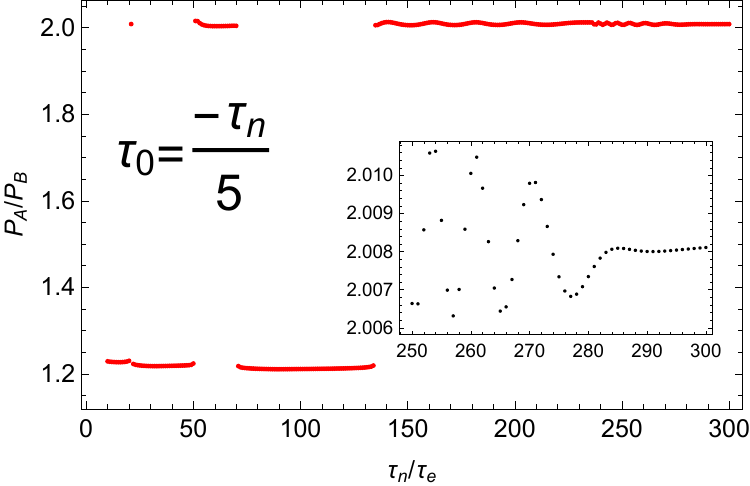}
               
        }~\subfigure[ ]{       
            \includegraphics[width=0.4\textwidth, height=0.25\textwidth]{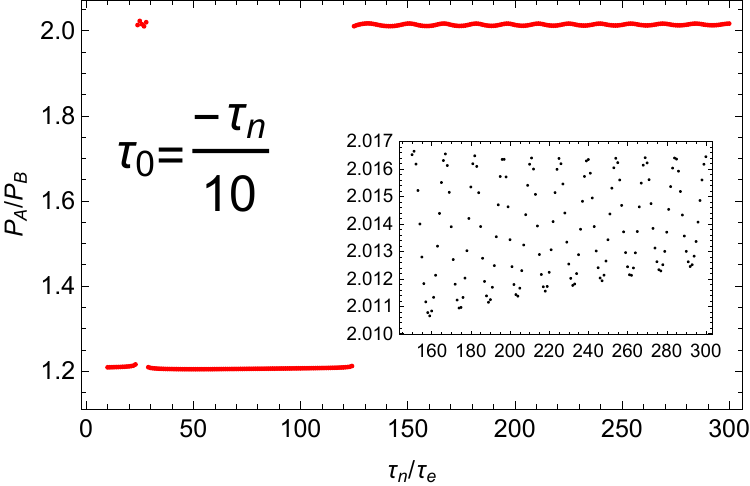}  
        }
         
    \caption{
     The end state at $n' t =5\times 10^4$ where the period ratios are extracted for various $\tau_0$. The inset for each plot provides a zoomed-in illustration for the finite amplitude libration regime and the possible fixed point regime.
     }
    \label{fig:array4}
\end{figure*}

With the results from Sec. \ref{sec:nomig} we can now discuss evolutions in the more general settings, with the migration torques turned on. In \citetalias{Goldreich:2013rma} with no $\tau_0$-related terms in the equation of motion, it is shown that there are three parameter regimes of interest. The system is trapped in resonance with fixed $Q$ and $e$ for
\begin{align}
\mu' > \frac{(m-1)}{\sqrt{3} m^{3/2} \beta_0} \left ( \frac{\tau_e}{\tau_n}\right )^{3/2}
\end{align}
and for 
\begin{align}\label{eq:mulow}
\mu' < \frac{(m-1)^2}{8\sqrt{3}m^{3/2}\beta_0} \left ( \frac{\tau_e}{\tau_n}\right )^{3/2}
\end{align}
the system is trapped in resonance in finite duration $\sim \tau_e$. When the resonance locking breaks down the system evolves with monotonically changing period ratios and decreasing eccentricities. Between  these two limits the system asymptotes a state with finite libration amplitude. Notice that the actual transition between resonance locking and rotational orbits in the phase space may significantly differ from the analytical criteria in Eq.~\eqref{eq:mulow} for small $\mu'$ and/or $\tau_n$, because the adiabatic approximation used to derive the analytical formulas  may break down \citepalias{Goldreich:2013rma}.

In Fig.~\ref{fig:array2} we present the evolution of a pair of planets according to Eq.~\eqref{eq:dne} but with the interfering density waves terms neglected. These figures essentially show the same qualitative features as Fig.~$5$ in \citetalias{Goldreich:2013rma}.
In the top row where $\tau_e =\tau_n/50$, the system exits the resonance within $\sim \tau_e$ timescale and follows with monotonic period ratios afterwards. Notice that although the period ratio is monotonically changing in this regime, we find that the resonant angle $Q$ is  bounded  (also see related discussion in Sec.~\ref{sec:nume}). This is because the contribution from $m n'-(m-1) n$ is canceled with $\dot{\varpi}_B$. It is also particularly interesting because if $Q$ stays bounded, the interfering density wave effect stays in operation even if the period ratio is out of resonance-locking, as discussed in Sec.~\ref{sec:nume}.  In the middle row with $\tau_e =\tau_n/100$, the system undergoes finite amplitude libration in the phase spacetime, so that both the eccentricity and period ratio oscillates around a fixed value. In the bottom row with $\tau_e = \tau_n/200$, the system asymptotes to a state with a fixed eccentricity $\sim 0.03$ and a period ratio offset $\sim 0.005$.

With the $\tau_0$ terms included, the evolution may be modified significantly. In Fig.~\ref{fig:array3} we present the case with the same parameters as in Fig.~\ref{fig:array2} but with $\tau_0=\pm\tau_n/10$.
For the positive $\tau_0$ cases, the system preferably lends on an asymptotic state with constant resonant angle and eccentricity (not shown in the plot). The period ratios are also asymptotically constant with $\le 0.3\%$ difference from $2$. On the other hand, for the negative $\tau_0$ case, the system no longer exhibits finite amplitude libration as shown in panel (c,d) of Fig.~\ref{fig:array2}, but instead loses the resonance locking without keeping bound $Q$. The period ratios are monotonically decreasing for the cases with $\tau_e =\tau_n/50$ and $\tau_e=\tau_n/100$, but oscillating around a fixed value for $\tau_e =\tau_n/200$. In the latter case the phase-space trajectory is qualitatively similar to the red trajectory in Fig.~\ref{fig:phase1}. Based on this particular system setup, it seems negative $\tau_0$ more likely drives the system far away from resonant period ratios, which is a phenomenon to be understood from the {\it Kepler} observations \citep{deck2015migration}.

%In the bottom row where we also find a stationary state, the eccentricity is damped with respect the value in Fig.~\ref{fig:array2} and the offset in period ratio from the exact resonance is higher. In the middle row with $\tau_e = \tau_n/100$, the system no longer exhibits finite amplitude libration as shown in Fig.~\ref{fig:array2}, but instead loses the resonance locking. In addition, in the top row with $\tau_e =\tau/50$, the system goes back to a stationary state with a fixed eccentricity and period ratio, as compared to the out-of-resonance behavior in  Fig.~\ref{fig:array2}. It is evident that the presence of interfering density wave terms can significantly modify the resonance dynamics of the  system.

In order to understand the ``end state" of the system with the influence of $\tau_0$ terms for a larger range of parameters, we perform a series of numerical evolutions using Eq.~\eqref{eq:dne} but with different $\tau_e$ and $\tau_0$. For each set of evolution we extract the period ratio at the end of the simulation, with $t_{\rm end} n' =5 \times 10^4$. In the top left panel of Fig.~\ref{fig:array4}, such an evolution is shown for $1/\tau_0 =0$. We find that roughly between $\tau_n/\tau_e \sim 180$ and $\tau_n/\tau_e \sim 70$ the ``end-state" period ratios show variations at the fixed end-state time, which is the consequence of finite-amplitude libration (i.e., the middle panel of Fig.~\ref{fig:array2}). With smaller $\tau_n/\tau_e$ than $70$  the period ratio is significantly smaller than 2, which corresponds to the non-resonant regime. On the other hand, for $\tau_n/\tau_e > 180$, the period ratio barely oscillates, which corresponds to the fixed point regime.

As the negative $\tau_0$ terms are included, the transition between the non-resonant regime and the libration regime shifts to lower $\tau_n/\tau_e$, as we can find for the cases with $\tau_0 =-\tau_n/10, -\tau_n/5$. 
%\textcolor{green}{Except for the isolated island in the very low $\tau_n/\tau_e$ regime, this transition between the non-resonant regime and the libration regime actually happens at large $\tau_n/\tau_e$, e.g., $\tau_n/\tau_e>100$}
%Similar shift also applies for the transition between the libration regime and the fixed-point regime.
In addition, as $|1/\tau_0|$ increases, a new non-resonant regime appears in the middle of the regime with roughly constant period ratios, as shown in panel (c,d) of Fig.~\ref{fig:array4}. %In fact, the system shown by the middle panel of Fig.~\ref{fig:array3} exactly resides in this new non-resonant regime.
The presence of additional non-resonant regimes is consistent with our earlier observation that negative $\tau_0$s tend to drive more system configurations away from the resonant period ratios. On the other hand, for positive $\tau_0$, i.e., $\tau_0=\tau_n/10$ as shown in the panel (b), most of the parameter range correspond to a system in resonance locking. Positive $\tau_0$ likely enhances the chance of resonance locking. In summary, the inclusion of $\tau_0$ terms in the evolution equations gives rise to more complex structures in the parameter phase space of such pairs of planets.

\section{Observational Implications}\label{sec:ob}

\begin{figure}
	\centering
    \includegraphics[clip=true, width=0.45\textwidth]{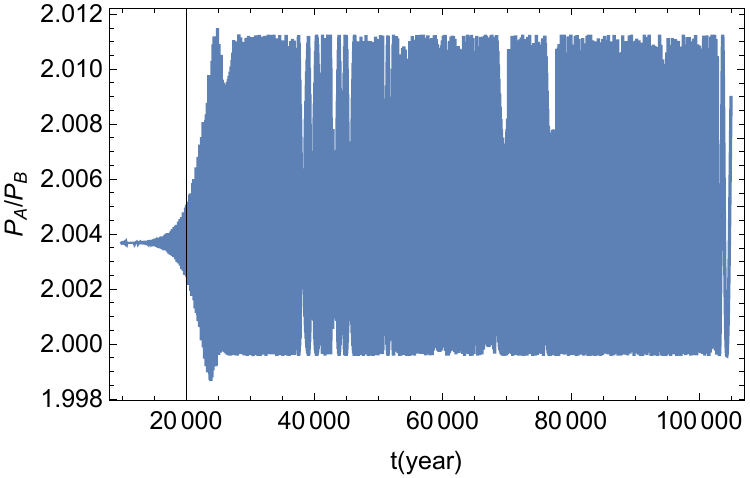}
	\caption{The period ratio of the evolution of a pair of planets with $\mu_B=0, \mu_A= 10^{-4}$ (with $\tau_0$ terms removed). The damping timescales are assumed to be $\tau_{nA}=10^5$ years and $\tau_{eB} =\tau_{nA}/100$. The value of $\tau_{eA}$ is not important as the eccentricity of object A is negligible in the quasi-stationary state. The system is initially released at $r_A =0.49 $AU and $r_B=0.3$ AU, with the star mass being $0.31 M_\odot$.} 
	\label{fig:pr1}
\end{figure}

\begin{figure}
	\centering
    \includegraphics[clip=true, width=0.45\textwidth]{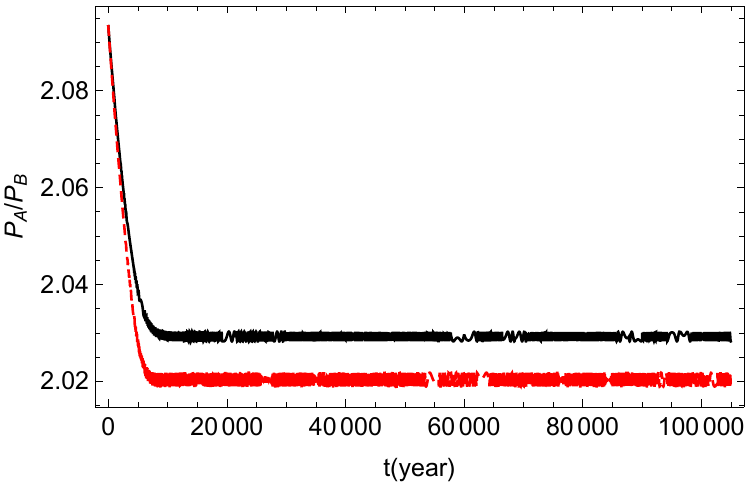}
	\caption{The evolution of a pair of planets with the same system parameters as in Fig.~\ref{fig:pr1}, except that the $\tau_0$ terms are turned on. For the red dashed line $\tau_0$ is set to be $-\tau_{nA}/20$ and for the black solid line $\tau_0$ is set to be $-\tau_{nA}/30$.} 
	\label{fig:pr2}
\end{figure}

The multi-planet systems discovered that by the {\it Kepler} mission have shown interesting observational signatures \citep{Fabrycky:2012rh}. First of all, most planets are found to reside away from MMRs. Secondly, for these found trapped in resonances, the period ratios are usually $1\%-2\%$ larger than the exact resonance. By introducing the damping terms in the eccentricity and semi-major axis, \citetalias{Goldreich:2013rma} manages to explain why it is rare to find planet-pairs trapped in resonances. However, their formalism suggests much smaller offsets (of period ratios) from the exact resonance for these pairs trapped in resonances.

In order to explain the period ratio offset, it was suggested that the tidal damping of eccentricity may play a role \citep{Lithwick:2012qt,Batygin:2012tu}, although \citet{Lee:2013pha} claims that tidal damping cannot account for the measured offsets with reasonable  tidal parameters. On the other hand, the work by \citet{Podlewska-Gaca:2011yrd,Baruteau:2013vv} suggests that the  density waves from the companion damped around  the planet may contribute to larger period ratios, which is possibly more efficient if a gap opens by the planet. In this section we investigate the effect coming from interfering density waves, without introducing extra dissipation mechanisms.

We extend the formalism discussed in Sec.~\ref{sec:mod} by including the dynamical evolution for both the inner and outer objects. The evolution equations are (we are dealing with a $2:1$ resonance) \citepalias{Goldreich:2013rma}
\begin{align}\label{eq:both}
\dot{n}_B &=1.89 \mu_A n^2_B (1.19 e_B \sin \phi_B -0.428 e_A \sin \phi_A) +\frac{n_B}{\tau_{nB}} +\frac{3 n_B e^2_B}{\tau_{eB}} \nonumber \\
& - \frac{6 e_B n_B \cos \phi_B}{\tau_0}, \nonumber \\
\dot{n}_A &=-6 \mu_B n^2_A (1.19 e_B \sin \phi_B -0.428 e_A \sin \phi_A) +\frac{n_A}{\tau_{nA}} +\frac{3 n_A e^2_A}{\tau_{eA}}\,, \nonumber \\
\dot{e}_B & = 0.75 \mu_A n_B \sin \phi_B -\frac{e_B}{\tau_{eB}} -\frac{\cos \phi_B}{\tau_0},\nonumber \\
\dot{e}_A &= - 0.428 \mu_B n_A \sin \phi_A -\frac{e_A}{\tau_{eA}},\nonumber \\
\dot{\varpi}_B & = -0.75 \mu_A n_B \frac{\cos \phi_B}{e_B},\nonumber \\
\dot{\varpi}_A & = 0.428 \mu_B n_A \frac{\cos \phi_A}{e_A}
\end{align}
with $\phi_{A,B} := \vartheta -\varpi_{A,B} : =2 \lambda_A-\lambda_B-\varpi_{A,B}$. Notice that in Eq.~\eqref{eq:both} we have only included the interfering density wave term $\tau_0$ for the evolution equations of object B for simplicity. This is a good approximation as later on we shall focus on the limit that $\mu_A \gg \mu_B$. In addition, the evolution equation for $\vartheta$ is 
\begin{align}\label{eq:theta}
\dot{\vartheta} = 2n_A-n_B\,,
\end{align}
which together with Eq.~\eqref{eq:both} provide seven evolution equations for seven variables $n_A, n_B, e_A, e_B , \varpi_A, \varpi_B, \vartheta$. Let us now consider the case with $\mu_A \gg \mu_B$ discussed in \citetalias{Goldreich:2013rma}. By setting $\mu_B=0, \mu_A =4 \times 10^{-4}$ and $\tau_{eB}/\tau_{nA}=0.022$, the period ratio can reach a level $\sim 1.5 \%$. However, most of the  {\it Kepler} pairs near the $2:1$ resonance has $\mu_A$ being much smaller than $4 \times 10^{-4}$, so that it was conjectured that additional eccentricity damping mechanism may be in operation.

In Fig.~\ref{fig:pr1} we are presenting an evolution using Eq.~\eqref{eq:both} and Eq.~\eqref{eq:theta} (but with $\tau_0$ terms removed) similar to Fig.~$12$ of \citetalias{Goldreich:2013rma}.The $\mu_A$ is assumed to be $10^{-4}$, which is a factor of four smaller than the value used in \citetalias{Goldreich:2013rma}, and $\tau_{eB}$ is set to be $\tau_{nA}/100$. In the stationary state the system undergoes large amplitude librations, with the average period ratio offsets from $2$ by about $0.3\%$, which is clearly much smaller than $\it Kepler$ observations. This is consistent with the findings in \citetalias{Goldreich:2013rma}, that $\mu_A$ has to be set with values higher than those observed in order to explain the $1\%-2\%$ period ratio offsets.

In Fig.~\ref{fig:pr2} we present the evolution of planet pairs with interfering density wave terms included. We find that with $\tau_0 \sim -\tau_{nA}/30$ is the ratio offset can achieve the $1.5\%$ level without the requirement of increasing the mass $\mu_A$. This $\tau_0$ corresponds to 
\begin{align}
\frac{1}{|\tau_0 \beta_0 \mu' n |} \approx 0.18\,,
\end{align}
so that the equivalent force coming from interfering density waves is about $20\%$ of the mutual gravitational interaction for the resonant dynamics. 
%With the definition of $\tau_0$ from Eq.~\eqref{eq:dtau0} and the scaling of $\tau_{nA}$ from Eq.~\eqref{eq:tauscaling}, it is reasonable to justify a factor of five difference between $\tau_0$ and $\tau_{nA}$. The additional factor may come from accounting for the vertical disc structure to evaluate the torque of interfering density waves and/or the undetermined numerical factors in the scaling laws in Eq.~\eqref{eq:tauscaling}.
The quasi-stationary part of Fig.~\ref{fig:pr2} can be understood as follows. This state is similar to the rotational state as depicted by the red and black trajectory in Fig.~\ref{fig:qcombine}, although here the rotation amplitude is stationary due to the contribution from both $\tau_0$ and $\tau_{eB}$ related terms. If we follow the expansion in Eq.~\eqref{eq:approx} for $n_B$ and $e_B$, we find that the unperturbed trajectory (without the $\tau_{eB}, \tau_{nA}, \tau_0$ terms) leads to
\begin{align}
\delta n_B &= -\frac{9 \mu_A e_{B0} n^2_{B0}}{4 \omega_0}\,,\nonumber \\
\delta e_{B} & = -\frac{3 \mu_A n_{B0}}{4 \omega_0}
\end{align}
where $\omega_0=2n_A-n_{B0}$. The corresponding conserved quantity $k$ for this system is given by
\begin{align}
k = \frac{3 e^2_B}{2} -\frac{3 \mu_A \cos \phi_B}{4 e_B} +\frac{\dot{Q}}{n_B}\,.
\end{align}
Now we treat the terms associated with $\tau_0, \tau_{eB}, \tau_{eA}$ as perturbations, and find that
\begin{align}
\dot{k} =3 e_B \dot{e}_B +\frac{2\dot{n}_A}{n_B}-\frac{\dot{n}_B}{n_B} =-\frac{6 e^2_B}{\tau_{eB}}+\frac{3 e_B \cos Q}{\tau_0}+\frac{1}{\tau_{nA}}\,.
\end{align}
The stationary state requires that the time average of $\dot{k}$ is zero, i.e., $\langle \dot{k} \rangle =0$, which  leads to
\begin{align}
\frac{1}{\tau_{nA}}-\frac{6 e^2_{B0}}{\tau_{eB}}-\frac{9 \mu_A n_{B0}}{4 \tau_0 \omega_0}\left ( 1+\frac{3 n_{B0}e^2_{B0}}{2\omega_0}\right )=0\,.
\end{align}
Dropping the subdominant term in the bracket we obtain the period ratio offset as
\begin{align}
\langle \frac{2 n_A}{n_B}-1 \rangle & =\frac{\omega_0}{n_{B0}} \approx \frac{9 \mu_A }{4  } \frac{\tau_{nA}}{\tau_0} \left ( 1-\frac{6 e^2_{B0} \tau_{nA}}{\tau_{eB}}\right )^{-1} 
\end{align}
which is approximately $-1.5\%$ for $\tau_0=-\tau_{nA}/30$ (in the asymptotic state $e_{B0} \sim 0.03$), agreeing well with the numerical evolution. It is evident that strong torque from the interfering density waves would lead to larger period offset. It is also possible that the interfering density waves provide the eccentricity damping mechanism to allow large period ratio offsets as observed in {\it Kepler} multi-planet systems. 
%At this stage we expect $\dot{e}_B \approx 0$ and
%\begin{align}
%0.75 \mu_A n_B \sin \phi_B \approx \frac{\cos \phi_B}{\tau_0}\,.
%\end{align}
%In addition, by noticing that $\phi_A \approx 0$ and $\dot{n}_B =2 %\dot{n}_A$ and using the first two lines of Eq.~\eqref{eq:both}, we find that
%\begin{align}
%e_B \approx -0.167 \frac{\tau_0}{\tau_{nA}}
%\end{align}
%which effectively explains the role of $\tau_0$ in damping the %eccentricity. As a result, the period ratio offset may be obtained by noticing that $\dot{\phi}_B \approx 0$ in the quasi-stationary state, which suggests that 
%\begin{align}
%\frac{2 n_A}{n_B}-1 &\approx -0.75 \frac{\mu_A}{e_B} \cos \phi_B \approx -4.5 \frac{\mu_A \tau_{nA}}{\tau_0}\,,\nonumber \\
%& = -1.3\% \times \left ( \frac{\mu_A}{10^{-4}} \right )\left (\frac{\tau_0} {\tau_{nA}/30} \right )^{-1}\,,
%\end{align}
%which justifies the findings shown in Fig.~\ref{fig:pr2}. It is possible that the interfering density waves provide the eccentricity damping mechanism to allow large period ratio offsets as observed in {\it Kepler} multi-planet systems. 

\section{Conclusion}\label{sec:con}

In this work we have discussed a new type of disc-mass interactions for a pair of point masses moving within an accretion disc. When the orbital phases of the masses are locked into a nearly constant resonant angle, 
the intefering density waves produce an extra piece of angular momentum flux that does not average to zero over orbital timescales. Using a two-dimensional theory with the same vertical averaging scheme in \citetalias{ward1988disk}, we compute the backreaction of the disc on to the planets near an MMR. We find that the backreaction torque is mainly contributed by the disc materials around the Lindbald resonance location close to one of the planets. The excitation amplitude $\Psi$ blows up for a standard two-dimensional theory but can be regularized by various smoothing schemes, e.g., the one discussed in \citetalias{ward1988disk}. For the particular $2:1$ inner resonance we compute the relevant torque acting on the inner planet due to this mechanism.
%The backreaction on the motion of masses produces distinctive features near the MMR. In the case that that the migration torques are neglected, the evolution of the pair of masses is no longer described by a Hamiltonian system, for which we find two asymptotic fixed points: one with constant resonant angle and decreasing eccentricity and the other with rotating resonant angle and growing eccentricity. With migration torques included, the new effect may still significantly modify the resonant dynamics - a system may switch from out-of-resonance state to a resonance-locking state with the assistance of interfering density waves. Note that  even when the resonance locking breaks down for the period ratio, the resonant angle may still stay at a constant level such that the density wave term still contributes significantly to the eccentricity evolution (c.f. Eq.~~\ref{eq:dne}). In many cases the eccentricity is further damped by the interfering density wave effect.

We have designed a set of hydrodynamical simulations to verify the analytical predictions. The simulations are also intrinsically two-dimensional, so that another smoothing scale $\epsilon$ is introduced in the gravitational potential following common practise in the literature of disc simulations. Interestingly, the analytical theory with this smoothing scheme predicts a different torque.  We perform a first simulation with a pair of planets following prescribed motion either in the 2:1 and 3:2 MMR state or the out-of-resonance state, and measure the corresponding disc gravitational force on the planets. We also perform separate hydrodynamical simulations with either the outer planet or inner planet alone, but with orbital parameters the same as in the first simulation , and measure the disc backreaction. In this way we can subtract the torques and obtain the additional torque due to the coupling between the inner planet and the density waves generated by the outer planet. We find that indeed the additional torque is mainly produced in the inner Lindblad resonance which is close to the orbit of the inner planet, with total value being $\sim -2.9\times 10^{-8}$ in code unit. Notice that the analytical theory  predicts a negative torque which is roughly consistent with the numerical values. Other signatures of the interference torque, i.e. the radius of location, the eccentricity and mass dependencies, the $Q$ dependence are all consistent with the analytical theory. Because the sign of the torque depends on the smoothing scheme as discussed earlier, it is clear that the actual torque sensitively depends on the detailed three-dimension gas structure in the co-rotation regime of the inner planet. In the future, we plan to perform a study in the three-dimensional scenario both with the analytical theory and numerical simulations to resolve the discrepancy. We have introduced an effective $\tau_0$ which can be both positive and negative (corresponding to positive and negative torques respectively) in the discussion of orbital dynamics in Sec.~\ref{sec:mod} and Sec.~\ref{sec:ob}.

With the new torque due to density wave interference, we have analyzed the dynamics of a part of planets within and outside of the resonance regime. We find that positive $\tau_0$ tend to drive more systems into the resonance regime, whereas negative $\tau_0$ more likely drive the systems away from the resonant period ratios. Indeed in {\it Kepler}'s observation most planet pairs are found to be away from the resonant period ratios. \citetalias{Goldreich:2013rma} proposed an explanation by introducing dissipation terms in the orbital evolution equations but \cite{deck2015migration} have pointed out that the model is insufficient to explain the data by considering more general mass ratios. It is therefore interesting to examine whether the interfering density wave effect can explain the large population of off-resonant pairs - an direction to explore in the future \citep{Ge2022}. On the other hand, we also find that negative $\tau_0$ can lead to a state with stable period ratio with $1\%-2\%$ offset  away from the exact resonant values. This is another observation signatures of planet pairs in {\it Kepler}'s data which can not be naturally explained in \citetalias{Goldreich:2013rma}. Here the interfering density waves provide an promising mechanism to produce relatively large period ratio offsets, but more extensive studies in the relevant parameter space are necessary to test whether it is fully consistent with data.

At this point, it may be interesting to generalize the effect due to interfering density waves to ``resonant dissipations" for systems under resonance as density wave emission is one form of dissipation in a resonant process. The essence of this effect is that dissipative mechanisms may dynamically depend on the resonant angle, so that they may introduce nontrivial influence on the resonant dynamics. For example,  one may imagine that the tide-driven migration in planet-satellite systems may exhibit similar phenomena. The planet tides excited by the satellites \citep{goldreich1965explanation} may coherently interfere with each other to produce additional resonant dissipation.
In  EMRI systems relevant for space-borne gravitational wave detection, with a single stellar-mass black hole orbiting around a massive black hole an orbital resonance may still arise because of the beating between different degrees of freedom of the orbit \citep{Flanagan:2010cd,Yang:2017aht,Bonga:2019ycj,Pan:2023wau}, which have different cyclic frequencies in the strong-gravity regime.  The main dissipation mechanism for these systems are gravitational wave radiation. The beating of gravitational waves of different harmonics near the resonance may give rise to an extra resonant dissipation  that depends on the resonant angle and modify the resonant dynamics in a nontrivial manner.

\section*{Acknowledgments}

We thank Houyi Sun for helpful discussions and the anonymous referee for many constructive comments.
 H. Y. is supported by the Natural Sciences and
Engineering Research Council of Canada and in part by
Perimeter Institute for Theoretical Physics.
 Research at Perimeter Institute is supported in part by the Government of Canada through the Department of Innovation, Science and Economic Development Canada and by the Province of Ontario through the Ministry of Colleges and Universities. 
Y.P.L. is supported in part by the Natural Science Foundation of China (grant NO. 12373070, and 12192223), and Natural Science Foundation of Shanghai (grant NO. 23ZR1473700).
The calculations have made use of the High Performance Computing
Resource in the Core Facility for Advanced Research Computing
at Shanghai Astronomical Observatory.

\section*{Data availability}
The data underlying this article will be shared on reasonable request to the corresponding author.

\bibliographystyle{mnras}
\bibliography{reference}
%
%%%%%%%%%%%%%%%%%%%%%%%%%%%%%%%%%%%%%%%%
%%%%%%%%%%%%%%%%%%%%%%%%%%%%%%%%%%%%%%%%
%%%%%%%%%%%%%%%%%%%%%%%%%%%%%%%%%%%%%%%%
\bsp
\label{lastpage}
%%%%%%%%%%%%%%%%%%%%%%%%%%%%%%%%%%%%%%%%
%%%%%%%%%%%%%%%%%%%%%%%%%%%%%%%%%%%%%%%%
\end{document}